\documentclass[fleqn,usenatbib]{mnras}

\usepackage{newtxtext,newtxmath}
\usepackage[T1]{fontenc}

\DeclareRobustCommand{\VAN}[3]{#2}
\let\VANthebibliography\thebibliography
\def\thebibliography{\DeclareRobustCommand{\VAN}[3]{##3}\VANthebibliography}

\usepackage{graphicx}	% Including figure files
\usepackage{amsmath}	% Advanced maths commands
\usepackage[normalem]{ulem}

\newcommand{\ramses}{{\rm \small RAMSES}\ }

\newcommand{\ramsesrt}{{\rm \small RAMSES-RT}\ }

\newcommand{\kmsec}{${\,\rm {km\,s^{-1}} }$}

\newcommand{\sigmag}{$\sigma_{\rm g}$\ }
\newcommand{\sigmagalt}{$\sigma_{\rm g}$}
\newcommand{\Msol}{\,{\rm M_\odot}}
\newcommand{\Msolyr}{\,{\rm M_\odot\,yr^{-1}}}

\newcommand{\Halpha}{{\rm H$\alpha$}\ }
\newcommand{\Halphaalt}{{\rm H$\alpha$}}
\newcommand{\Htwo}{{\rm H$_2$}\ }
\newcommand{\HII}{H{\small II}\ }
\newcommand{\HI}{H{\small I}\ }

\newcommand{\emHa}{$\epsilon_{\rm H\alpha}$\ }
\newcommand{\emHaalt}{$\epsilon_{\rm H\alpha}$}
\newcommand{\LHa}{L$_{\rm H\alpha}$\ }
\newcommand{\LHaalt}{L$_{\rm H\alpha}$}

\newcommand{\sigmaHa}{$\sigma_{\rm H\alpha}$\ }
\newcommand{\sigmaHaalt}{$\sigma_{\rm H\alpha}$}

\title[]{The origin of the H$\alpha$ line profiles in simulated disc galaxies} %  withint the multiphase

\author[T. Ejdetj{\"a}rn et al.]{
Timmy Ejdetj{\"a}rn$^{1}$\thanks{E-mail: timmy.ejdetjarn@astro.su.se}, Oscar Agertz$^2$, G{\"o}ran {\"O}stlin$^1$, Martin P. Rey$^3$, Florent Renaud$^{4,5}$
\\
$^{1}$Oskar Klein Centre, Department of Astronomy, Stockholm University, 106 91 Stockholm, Sweden\\
$^{2}$Lund Observatory, Division of Astrophysics, Department of Physics, Lund University, Box 43, SE-221 00 Lund, Sweden\\
$^{3}$Sub-department of Astrophysics, University of Oxford, DWB, Keble Road, Oxford, OX1 3RH, UK\\
$^{4}$Observatoire Astronomique de Strasbourg, Universit{\'e} de Strasbourg, CNRS UMR 7550, F-67000 Strasbourg, France\\
$^{5}$University of Strasbourg Institute for Advanced Study, 5 all{\'e}e du G{\'e}n{\'e}ral Rouvillois, F-67083 Strasbourg, France
}

\date{Accepted XXX. Received YYY; in original form ZZZ}

\pubyear{2024}

\begin{document}
\label{firstpage}
\pagerange{\pageref{firstpage}--\pageref{lastpage}}
\maketitle
\begin{abstract}
Observations of ionised H$\alpha$ gas in high-redshift disc galaxies have ubiquitously found significant line broadening, $\sigma_{\rm H\alpha}\sim 10-100\,{\rm km\,s^{-1}}$. To understand whether this broadening reflects gas turbulence within the interstellar medium (ISM) of galactic discs, or arises from out-of-plane emission in mass-loaded outflows, we perform radiation hydrodynamic (RHD) simulations of isolated Milky Way-mass disc galaxies in a gas-poor (low-redshift) and gas rich (high-redshift) condition and create mock H$\alpha$ emission line profiles. We find that the majority of the total (integrated) ${\rm H\alpha}$ emission is confined within the ISM, with extraplanar gas contributing $\sim45\%$ of the extended profile wings ($v_z\geq 200$\kmsec) in the gas-rich galaxy. This substantiates using the ${\rm H\alpha}$ emission line as a tracer of mid-plane disc dynamics. We investigate the relative contribution of diffuse and dense ${\rm H\alpha}$ emitting gas, corresponding to diffuse ionised gas (DIG; $\rho \lesssim 0.1\,{\rm cm^{-3}}$, $T\sim 8\,000$\,K) and H{\small II} regions ($\rho \gtrsim 10\,{\rm cm^{-3}}$, $T\sim 10\,000$\,K), respectively, and find that DIG contributes $f_{\rm DIG}\lesssim 10\,\%$ of the total L$_{\rm H\alpha}$. However, the DIG can reach upwards of $\sigma_{\rm H\alpha} \sim 60-80\,{\rm km\,s^{-1}}$ while the H{\small II} regions are much less turbulent $\sigma_{\rm H\alpha}\sim10-40\,{\rm km\,s^{-1}}$. This implies that the $\sigma_{\rm H\alpha}$ observed using the full ${\rm H\alpha}$ emission line is dependent on the relative ${\rm H\alpha}$ contribution from DIG/H{\small II} regions and a larger $f_{\rm DIG}$ would shift $\sigma_{\rm H\alpha}$ to higher values. Finally, we show that $\sigma_{\rm H\alpha}$ evolves, in both the DIG and H{\small II} regions, with the galaxy gas fraction. Our high-redshift equivalent galaxy is roughly twice as turbulent, except for in the DIG which has a more shallow evolution.
% Outflows and extraplanar gas contribute a very small portion of the total \Halpha emission line, but this depends on the strength of galaxtic outflows and the impact of \Halpha dust scattering.

\end{abstract}


%%%%%%%%%%%%%%%%%%%%%%%%%%%%%%%%%%%%%%%%%%%%%%%%%%

%%%%%%%%%%%%%%%%% BODY OF PAPER %%%%%%%%%%%%%%%%%%

%%%%%%%%%%%%%%%%%%%%%%%%%%%%%%%%%%%%%%%%%%%%%%%%%%%%%%%%%%%%%%%%%%%%%%%%%%%%%%%%%%%%%%%%%%%%%%
\section{Introduction}\label{sec:intro}
% Gas within the interstellar medium (ISM) of disc galaxies is known to be supersonically turbulent in 

The gas inside galactic discs, the interstellar medium (ISM), consist of various gas phases spanning several order of magnitude of densities and temperatures, from dense molecular clouds, where stars are formed, to warm ionised gas, which makes up a majority of the total gas volume but only a fraction of the mass \citep[][]{McKeeOstriker77}. In particular, the ionised gas phase is prominent in the disc, outflows and extraplanar gas \citep[][]{MacLowKlessen04, Haffner+09}{}{}, and has several formation channels through stellar and active galactic nuclei (AGN) feedback \citep[e.g.][]{MacLowKlessen04,Haffner+09, HeckmanBest14, Somerville+15}{}{}. Observations of ionised gas thus trace the complex dynamics caused by a variety of feedback effects, which might not represent the bulk motions or overall dynamical state of the galaxy. %; while molecular gas is primarily found within the plane of the disc.

% Observationally, emission lines trace a specific gas phase and, thus, ionised gas tracers represent the dynamics associated with the formation of ionised gas, which might not represent the bulk motions or properties of the galaxy.

% The reprocessing of ionised, neutral, and molecular gas in galaxies is complex, owing to the various heating and cooling mechanisms, but the dynamics and physical properties of each phase are commonly derived through emission lines that trace a specific gas phase. Due to the variation in nature between

%reprocessing/transitioning of material to different gas phases
%emission lines primarily trace one gas phase, so dynamics varies with this choice
% and to understand this better observers rely on emission lines that trace specific gas tracers.
% The nature and formation of these gas phases is not fully understood and is relevant for the reprocessing of gas between phases, but also for observations which use emission lines that trace ionised gas in order to interpret the dynamics and physical state of galaxies. 

A prominent difference between gas phases is their kinematics. Gas turbulence, often quantified as the velocity dispersion \sigmagalt, is a fundamental property of galactic discs, with supersonic motions driving crucial physical processes such as star formation \citep[][]{Larson81, MacLowKlessen04, Renaud+12, Federrath2018}{}{}, metal mixing \citep[][]{YangKrumholz12,ArmillottaKrumholz18}{}{}, and density distributions \citep[][]{McKeeOstriker07}{}{}. The ionised gas in local galaxies has been observed to be more turbulent, \sigmag$\sim15-30$\kmsec, \citep[][]{Epinat+10, Moiseev+15, Green+14, Varidel+16, Law+22}{}{} than both neutral \citep[][]{Tamburro+09,Wisnioski+11}{}{} and molecular gas \citep[][]{Nguyen-Luong+16, Levy+18}{}{}, \sigmag$\sim10$\kmsec. Importantly, the thermal energy difference is not enough to explain the offset in turbulence between these gas phases. The driver of this supersonic, multi-phase turbulence is not fully understood \citep[see the reviews by][]{ElmegreenScalo04, MacLowKlessen04, Glazebrook13}{}{}, but it is believed that a combination of stellar feedback and gravitational instability could inject a majority of the turbulent energy \citep[e.g.][]{Dib06, Agertz+09, Faucher-Giguere+11, HaywardHopkins17, Krumholz+18, Orr+20,Ejdetjarn+22,Forbes+23}{}{}.

%Locally, the ionised gas in disc galaxies is much less turbulent than at higher redshift 
The problem is further compounded, with velocity dispersions being reported much higher at higher redshift ($z\sim2$), where line widths on the order of \sigmag$\sim 30-100$\kmsec\ are observed \citep[][]{Law+09, Cresci+09, Genzel+11, Alcorn+18,Ubler+19,Girard+21,Lelli+23}{}{}. However, high redshift observations suffer from poor spatial resolutions, which increases the risk of beam smearing and, in combination with uncertain galaxy inclination corrections, can raise the velocity dispersion by a factor of 3-5 \citep[see e.g.][]{DiTeodoroFraternali15, Rizzo+22, Ejdetjarn+22, Kohandel+23}. Additionally, insufficient spectral resolution can also artificially broaden line profiles \citep[see e.g.][]{Lelli+23}{}{} and unresolved, or barely resolved, lines have an upper limit for \sigmag set by the instrument's resolution.

Furthermore, galaxies undergoing interactions or mergers, e.g. local (Ultra) Luminous Infrared Galaxies \citep[][]{Armus+87, Sanders+88, Carpineti+15}{}{}, have been observed to feature very high levels of turbulence, with broad emission lines ($\sigma_{\rm g} \gtrsim 100\,{\rm km\,s^{-1}}$) even in the cold molecular gas phase \citep[][]{Veilleux+09, Renaud+14, Scoville+15, Scoville+17}{}{}. Similarly, ionised gas in local low-mass starburst galaxies, which are dispersion-dominated and likely interacting, have been found with $\sigma_{\rm g}\sim30-100$\kmsec \citep[e.g.][]{Ostlin+01, Green+14, Herenz+16, Bik+22}{}{}. Disentangling the complex gas kinematics that arise during galaxy interactions from internal \& extraplanar motions (i.e. within the disc and in/outflows), and its impact on \sigmagalt, is a difficult task, especially for high-redshift observations with low spatial and spectral resolution.

One of the most common emission lines used to infer properties of warm ionised gas in both local and high-redshift galaxies is the \Halpha line, due to its strength, ubiquity at all redshifts, and straight-forward dust correction through the Balmer decrement. The largest \Halpha flux within galaxies comes from \HII regions, which are the warm $T\sim 10^4$\,K, dense $n_{\rm e}\sim 10-10^3\,{\rm cm^{-3}}$ gas regions with recent star formation, and are almost fully ionised due to the ionising flux from massive OB stars \citep[e.g.][]{OsterbrockFerland06}{}{}. But several studies have found that diffuse ionised gas (DIG; also referred to as the warm ionised medium, WIM), with densities $n_{\rm e}\lesssim 0.08\,{\rm cm^{-3}}$ and temperatures $T\sim8000$\,K \citep[see][for a review]{Haffner+09}{}{}, may be the dominant contributor ($f_{\rm DIG}\gtrsim50\,\%$) towards to total \Halpha luminosity in local galaxies \citep[e.g.][]{Ferguson+96, Hoopes+96, Oey+07, Kreckel+16, Belfiore+22}{}{}. However, this conclusion is debated, given the large spread of $f_{\rm DIG}$ with values as low as $f_{\rm DIG}\sim 10\%$ in some galaxies \citep[e.g.][]{Oey+07, Blanc+09, DellaBruna+22, Micheva+22}{}{}. This adds more complexity when trying to discern the underlying origin of the \Halpha line.
% , which is important to account for the ionisation budget in galaxies

Furthermore, the DIG and \HII regions inhabit different environments and are ionised by different mechanisms. A large fraction of the ionisation for the DIG is believed to come from leaking \HII regions \citep[][]{Haffner+09, Weilbacker+18}{}{} but many unrelated mechanisms likely contribute, e.g. supernovae shocks \citep[][]{CollinsRand01}{}{}, cosmic rays \citep[][]{Vandenbroucke+18}{}{}, and dissipation of turbulence \citep[][]{MinterSpangler97, Binette+09}{}{}. In the scenario that the mid-planar DIG originates from ionising photons leaking from \HII regions, it is possible these phases would share kinematics. However, observations of \Halpha in low-redshift galaxies indicate that \HII regions have $\sigma_{\rm H\alpha}\sim10-20$\kmsec while DIG reaches upwards of $\sigma_{\rm H\alpha}\sim 50$\kmsec\ \citep[][]{DellaBruna+20, denBrok+20, Law+22}{}{}. The distinct kinematic separation between these phases may lead to a spread in the observed \sigmaHa in galaxies, as the \Halpha profile might be dominated either by \HII regions or DIG. Disentangling the impact of these environments requires either highly resolved analysis of the gas properties or detailed profile-fitting of the different components that make up the \Halpha line. 

%  The necessary gas transport process in this paradigm connects to the study of galactic winds, which are defined as gas flows with a velocity component that exceeds the escape velocity of the galactic potential, and galactic outflows, which remain bound to their host galaxy.
Outflows are a natural part of star-forming disc galaxies, as stellar and AGN feedback effects continuously eject warm \& hot gas into the halo which can eventually cool and reaccrete onto the disc (known as a galactic fountain), fueling new stars. Strong outflows that exceed the escape velocity of the galaxy, referred to as galactic winds, are commonly associated with galaxies undergoing starbursts \citep[see][for reviews]{HeckmanBest14, Somerville+15}{}{}. This forms a complex dynamic structure above/below the disc which is multi-phase in nature, metal-rich, and with a wide dynamic range. The majority of extraplanar ionised gas is believed to come from extended DIG (eDIG) a few kpc from the disc \citep[e.g.][]{Rand96,Rossa+04,Haffner+09,Ho+16}{}{}. This gas exists either as a semi-static corona (perpetually heated by ionising photons from massive stars in the disc or runaway stars) or as part of ongoing outflows \citep[][]{Zurita+02, Haffner+09, Barnes+15, Dirks+23}{}{}. This suggests that the DIG, at least partially, inhabits an environment untouched by \HII regions and, thus, contributes the majority of extraplanar \Halpha emission.
%% Two key drivers of outflows suggested in literature: stellar feedback (primarily supernova) and radio jets from AGN \citep[see][for reviews]{HeckmanBest14, Somerville+15}{}{}.

Numerical simulations that capture the multi-phase nature of the ISM confirm that the warm, ionised gas is more turbulent than the colder, neutral or molecular gas phases \citep[][]{Orr+20, Ejdetjarn+22, Rathjen+23, Kohandel+23}{}{}. To accurately capture the ionised gas structure in galaxies, in particular the dense \HII regions, very high spatial resolution with an explicit treatment of radiative transfer effects is required \citep[see e.g.][]{Deng+23}{}{}. Recent radiation hydrodynamic (RHD) simulations of disc galaxies environments (entire discs or ISM patches) have investigated the \Halpha emission \citep[e.g.][]{Katz+19, Tacchella2022, Kohandel+23}{}{}, with a few RHD simulations investigating the nature of DIG and/or \HII regions \citep[][]{VandenbrouckeWood19, Tacchella2022, McClymont+24}{}{}. Notably, \citet[][see also \citealt{McClymont+24}]{Tacchella2022}{}{} simulated an isolated Milky Way-like galaxy and performed mock observations of the \Halpha emission with post-processing ray tracing. They decomposed the contribution of DIG and \HII regions to the \Halpha luminosity, finding an agreement with observations $f_{\rm DIG}\sim50\,\%$. However, no RHD simulation has yet to investigate the kinematic offset between the two gas phases that has been observed with the \Halpha line.

In this paper, we study how \Halpha emission from DIG and \HII regions changes as we go from a galaxy similar to the Milky Way today, to more gas-rich, rapidly star forming galaxies typical of high redshift ($z\sim 1-3$). Using hydrodynamical simulations of entire disc galaxies, with an explicit treatment of multi-frequency radiative transfer, we study to what degree extraplanar gas, as opposed to the disc ISM, contributes to \Halpha line profiles in such galaxies. 

This paper is organised as follows. In Section 2, we outline the setup of our models and the calculation of relevant parameters, such as the intrinsic velocity dispersion of the gas \sigmagalt. In Section 3, we visualise the simulations and create mock \Halpha line profiles to highlight strong \Halpha emitting regions within and outside the disc. Furthermore, we show how the intrinsic velocity dispersion of the ionised gas, traced by \Halphaalt, \sigmaHa changes between environments (i.e. dense and diffuse gas regions). In Section 4, we quantify contribution of outflows and extraplanar gas to the \Halpha mock line profiles and \sigmaHaalt. Finally, in Section 5 we summarise our conclusions.

\begin{table*}
    \centering 
    \caption{Properties of the simulated galaxies formed in this work. These values are calculated 200\,Myr after the last refinement step of each galaxy. The stellar and gas radius are exponential fits of the discs, which are slightly different than the initial radius due to the disc settling into a stable state and stellar feedback reshaping the disc. See Section~\ref{sec:ics} for specifics about the initial conditions and how values in this table were calculated. }
    \begin{tabular}{l l l l l l l l l}
        \hline\hline
        Name of run & $f_{\rm g}$ (\%) &  SFR ($\Msolyr$) & $M_*$ [$10^{10}\,{\rm M_\odot}$] & $M_{\rm g}$ [$10^{10}\,{\rm M_\odot}$] &  $r_{\rm *}$ [kpc] & $h_{\rm *}$ [kpc] &  $r_{\rm g}$ [kpc] & $h_{\rm g}$ [kpc] \\
        \hline 
        \texttt{fg10}   & 9  & 2  & 4.10 & 0.40 & 3.96 &  0.19 & 3.96 & 0.21 \\
        \texttt{fg50}   & 33 & 39 & 3.02 & 1.48  & 2.67 & 0.33 & 4.52 & 0.33 \\
        \hline
    \end{tabular}
    %% Fitting an exponential to fg10: r_d = 3.96 kpc, h_d = 0.19 kpc. fg50: r_d = 2.67 kpc (but gas has ~4.52 kpc), h_d = 0.33 kpc
    \label{tab:ICs}
\end{table*}

\begin{table*}
    \centering  
    \caption{ The \Halpha properties of the simulated galaxies calculated 200\,Myr after the last refinement step, as described in Section~\ref{sec:calculations}. } 
    \begin{tabular}{l l l l l l l l l}
        \hline\hline
        Name of run & L$_{\rm H\alpha}\ [10^{41}$ erg\,s$^{-1}$\,cm$^{-3}$] & $\sigma_{\rm H\alpha}$ [km\,s$^{-1}$]  & $\sigma_{\rm H\alpha,\,DIG}$ [km\,s$^{-1}$] & $\sigma_{\rm H\alpha,\,HII}$ [km\,s$^{-1}$] \\
        \hline 
        \texttt{fg10}  & $4.5\pm1$ &  $15\pm2$ & $43\pm4$ & $10\pm1$ \\
        \texttt{fg50}  & $45\pm5$ &  $35\pm5$ & $54\pm5$ & $29\pm3$ \\
        \hline
    \end{tabular}
    \label{tab:Halpha_values}
\end{table*}

\begin{figure*}
	\includegraphics[width=0.8\textwidth]{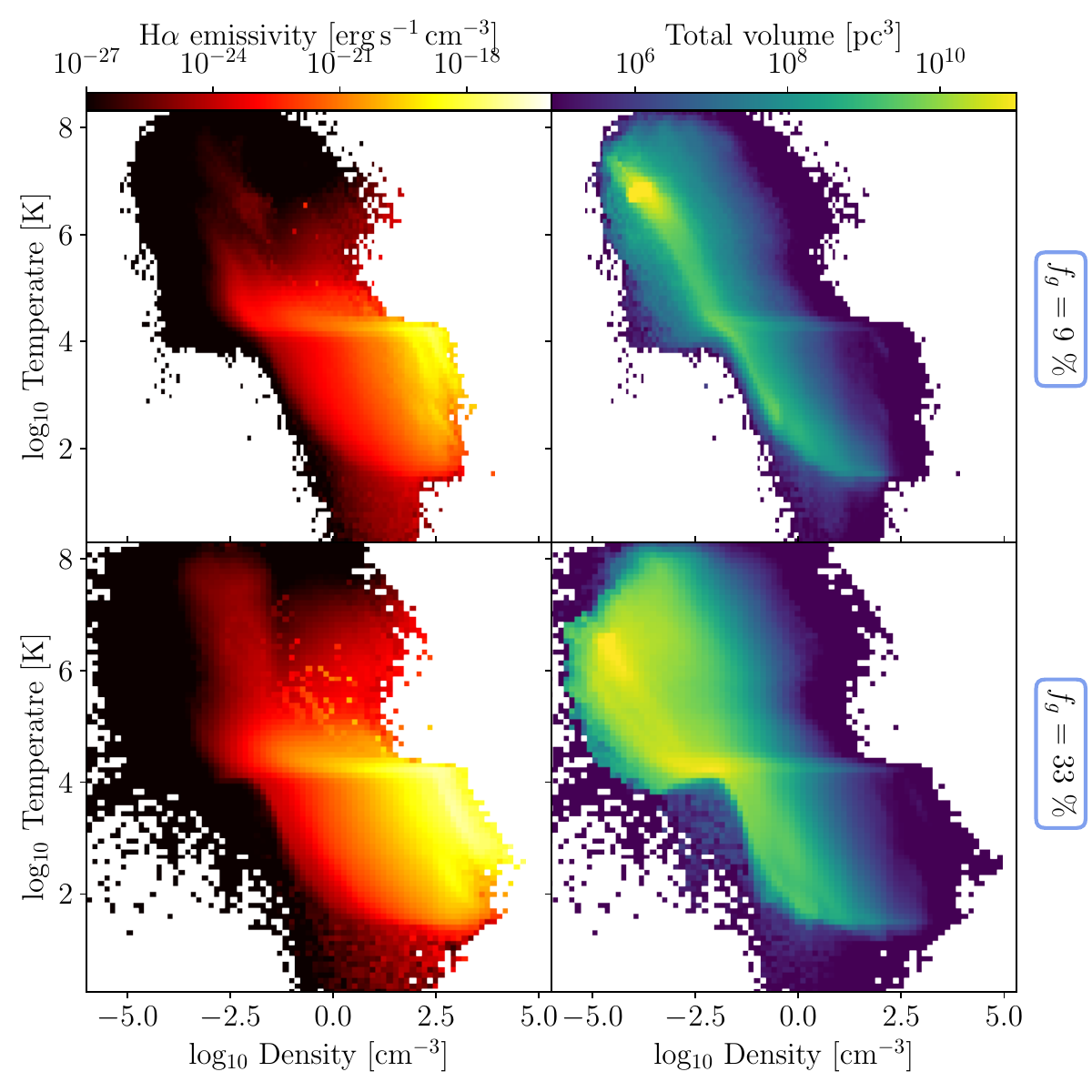}
    \caption{Colour diagrams of gas density-temperature inside the simulated galaxies. Each pixel represents all of the cells with those particular densities and temperatures. Left plots show the \Halpha emissivity weighted-mean of each cell while the right plots are coloured after the volume of cells in each pixel.}
    \label{fig:Halpha_phase}
\end{figure*}

%%%%%%%%%%%%%%%%%%%%%%%%%%%%%%%%%%%%%%%%%%%%%%%%%%%%%%%%%%%%%%%%%%%%%%%%%%%%%%%%%%%%%%%%%%%%%%
\section{Numerical method}
The simulations presented in this work were made with the radiation hydrodynamics (RHD) code \ramsesrt \citep[][]{Rosdahl+13, RosdahlTeyssier15}{}{}, which is a module of the $N$-body and hydrodynamical Adaptive Mesh Refinement (AMR) code \ramses \citep[][]{Teyssier02}. The code uses an HLLC Riemann solver \citep[][]{ToroSpruceSpeares94}{}{} to solve the Euler fluid equations for the dynamical interaction of gas with dark matter and stellar populations, using a second-order Godunov scheme, assuming an ideal mono-atomic gas with adiabatic index $\gamma = 5/3$. Gas interacts through gravity, hydrodynamics, radiation transfer (RT), and non-equilibrium cooling/heating. Dark matter and stars are represented as collisionless particles, their dynamics solved with a particle-mesh solver and cloud-in-cell interpolation \citep[][]{GuilletTeyssier11}{}{}.

In the coming sections, we outline the most relevant of the physics routines within our simulations, but refer to \citet[][]{Agertz+13}{}{} for a more detailed description of the feedback routines, and \citet[][see also \citealt{RosdahlTeyssier15}]{Rosdahl+13}{}{} for the RT routines \citep[see also][for details on how RT is applied in our simulations]{Agertz+20}{}{}.

%%%%%%%%%%%%%%%%%%%%%%%%%%%%%%%%%%%%%%%%%%%%%%%%%%%%%%%%%%%%%%%%%%%%%%%%%%%%%%%%%%%%%%%%%%%%%%
\subsection{Star formation and feedback physics}
In our simulations, star formation events are treated as stochastic events with star particles of masses $10^3\,{\rm M_\odot}$. Each particle represents a single-age stellar population with a  \citet[][]{Chabrier03}{}{} initial mass function. The number of star particles formed are sampled from a discrete Poisson distribution. Star formation follows a Schmidt law \citep[][]{Schmidt59, Kennicutt98} of the form
\begin{align}
    \dot\rho_* = \epsilon_{\rm ff} \frac{\rho_{\rm g}}{t_{\rm ff}},{\rm\ for\ }\rho_{\rm g}> \rho_{\rm *}.
\end{align}
The gas density threshold for star formation is chosen to be $\rho_{\rm *} = 100\,m{\rm _H\, cm^{-3}}$ \citep[above which most of the gas is molecular, see][and references therein]{GnedinTassisKravtsov09}{}{}. The free-fall time is $t_{\rm ff} = \sqrt{3\pi/32 G\rho}$ and the star formation efficiency per free-fall time is adopted as $\epsilon_{\rm ff} = 10\,\%$. This choice of $\epsilon_{\rm ff}$ has been shown to reproduce the observed low efficiency $\epsilon_{\rm ff} \sim 1\,\%$ inside giant molecular clouds \citep[][]{KrumholzTan07}{}{} and averaged over kpc scales \citep[][]{Bigiel+08}{}{}, when paired with a realistic stellar feedback treatment \citep[][]{AgertzKravtsov16, Grisdale2018, Grisdale2019}{}{}.  

We track the injection of momentum, energy, and metals of supernovae Type Ia and Type II, and stellar winds through subgrid recipes \citep[see][]{Agertz+13, AgertzKravtsov15}{}{}, including the technique of \citet[][]{KimOstriker15}{}{} for capturing the terminal momentum injected by individual supernova explosions \citep[for implementation details, see][]{Agertz21}. The radiation feedback is self-consistently treated by the RT extension, see Section~\ref{sec:RT}. Metal abundances are advected as scalars with the gas, for which we track iron and oxygen separately. These metallicities are combined, assuming a relative solar abundance \citep[][]{Asplund2009}{}{}, into a total metallicity, which is used for cooling, heating, and RT.

%%%%%%%%%%%%%%%%%%%%%%%%%%%%%%%%%%%%%%%%%%%%%%%%%%%%%%%%%%%%%%%%%%%%%%%%%%%%%%%%%%%%%%%%%%%%%%
\subsection{Radiative transfer}\label{sec:RT}

Stellar radiation is handled with moment-based radiative transfer using the M1 closure approach for the Eddington tensor \citep[][]{Levermore84}{}{}. This allows for RT to be described as a set of conservation laws and radiation can then be diffused via the AMR grid. Intercell flux is calculated with the Global Lax-Friedrichs function \citep[see][]{Rosdahl+13}{}{}. In order to avoid long computational times due to light propagation, we adopt a reduced speed of light approximation, $\bar c = 0.01\,c$ \citep[][]{Rosdahl+13}{}{}. The spectral energy distributions for stellar particles are taken from \citet[][]{BruzualCharlot03}{}{}, assuming single stellar population. Radiation is emitted in photon groups with a set range of frequencies, chosen to represent crucial physical processes. We adopt six photon groups that represent the infrared, optical, ionising radiation of H{\small I}, He{\small I}, He{\small II}, and photo-disassociation of molecular gas \citep[see][ for details]{Agertz+20}{}{}. %% (Lyman-Werner photons) 

The non-equilibrium chemistry and radiative cooling is solved for five ion species, neutral \& ionised hydrogen and helium, and we track their fractions ($x_{\rm H \small I }$, $x_{\rm H \small II }$,  $x_{\rm He \small I }$, $x_{\rm He \small II }$, $x_{\rm He \small III }$) along with the photon fluxes in every cell. For gas $T>10^4$\,K, metal cooling uses tabulated cooling rates from {\small CLOUDY} \citep[][]{Ferland+98}{}{} and $T\leq 10^4$\,K gas use the fine structure cooling rates from \citet[][]{RosenBregman95}{}{}. Molecular hydrogen \Htwo is modelled following \citet[][]{Nickerson+19}{}{} and accounts for formation, advection, destruction, and cooling. Self-shielding of the gas against ionising radiation is treated self-consistently by the RT module.

%%%%%%%%%%%%%%%%%%%%%%%%%%%%%%%%%%%%%%%%%%%%%%%%%%%%%%%%%%%%%%%%%%%%%%%%%%%%%%%%%%%%%%%%%%%%%%
\subsection{Simulation suite}\label{sec:ics}
The simulation data used in our analysis are isolated disc galaxies constructed to be Milky Way-like galaxies, in terms of their mass and size, but at different redshifts. By changing the gas fraction between simulations, each galaxy represents a Milky Way-like galaxy at a different stage of its evolution. The initial conditions used here are the same as in \citet[][hereafter Paper I]{Ejdetjarn+22}{}{} and are based on the isolated discs part of the {\small AGORA} project \citep[][]{Kim+2014, Kim+2016}{}{}. Briefly, the dark matter halo follows a NFW-profile \citep[][]{NavarroFrenkWhite96}{}{} with virial mass $M_{200}=1.1\cdot 10^{12}\,{\rm M_\odot}$ within $R_{200} = 205$\,kpc, and a dark matter concentration of $c=10$. The baryonic disc has a mass of $M_{\rm d}=4.5\cdot10^{10}\,{\rm M_\odot}$ and is represented with an exponential profile with scale length $r_{\rm d}=3.4$\,kpc and scale height $h=0.1\,r_{\rm d}$. The scale lengths are the same for both stellar and gas component, and the mass is divided according to the galaxy gas fraction (see paragraph below). The stellar bulge has a Hernquist-profile \citep[][]{Hernquist1990}{}{} with scale radius $r_{\rm b} = 0.1\,r_{\rm d}$. The dark matter halo and stellar disc are represented by $10^6$ particles and the bulge by $10^5$ particles. The galaxy is surrounded by a circumgalactic medium with density $\rho=10^{-5}\,{\rm m_H\, cm^{-3}}$ and with an ultraviolet background following \citet[][]{Faucher-Giguere+09}{}{}

We model the galaxy at different redshifts by changing the gas fraction $f_{\rm g} = M_{\rm g}/(M_{\rm g} + M_*)$ of the galaxy, keeping the total mass the same. The \texttt{fg50} galaxy corresponds to the Milky Way at $z\sim1.5$ \citep[based on SFR matching for Milky Way progenitor galaxies;][]{vanDokkum+13}{}{}, with initial gas fraction $f_{\rm g}=50\%$, which results in an initial stellar mass $M_*=2.25\cdot10^{10}\,{\rm M_\odot}$. The \texttt{fg10} galaxy is the Milky Way at $z\sim0$ and started with $f_{\rm g}=10\%$, thus having an initial stellar mass $M_*=4.05\cdot10^{10}\,{\rm M_\odot}$. The galaxies were initialised at coarser spatial resolutions for 100-200 Myr, to allow the gas and particles to reach an equilibrium, and subsequently refined to a maximum spatial resolution $\Delta x = 6$\,pc. We analyse the galaxies 150-200\,Myr after the last refinement step. The time between simulation outputs is 5 Myr and we analyse 10 outputs from each gas fraction. Thus, the calculated properties are time-averaged over 50 Myr unless otherwise stated. General properties of each galaxy at the time of analysis is presented in Table~\ref{tab:ICs}.

%%%%%%%%%%%%%%%%%%%%%%%%%%%%%%%%%%%%%%%%%%%%%%%%%%%%%%%%%%%%%%%%%%%%%%%%%%%%%%%%%%%%%%%%%%%%%%
\subsection{Calculating H$\alpha$ kinematic properties}\label{sec:calculations}
In this section we detail the calculations of various properties for the gas traced by the \Halpha emission line. We calculate the \Halpha signal directly from the gas in each cell by applying analytical equations of the recombination and collisional rate of the \Halpha emissivity \emHa in each simulation cell. The details of how these equations are derived and calculated is described in Appendix C of Paper I\footnote{We note that by using the energy of the \Halpha transition in their Eq. C2, it is indirectly and incorrectly assumed that the majority of \HI is constantly in the excited 2s state, rather than emitting back to the ground state. This can be easily corrected for by switching $hv$ in this equation with $\Delta E = 13.6\,{\rm eV}(1^2 - \ell_{\rm max}^{-2}) = 12.1\,{\rm eV}$. As collisional excitation contributes around 10-20\% to the total \Halpha signal \citep[$\leq10\%$ in the case of][]{Tacchella2022}{}{}, this does not affect the results in Paper I.}. This approach does not account for absorption and scattering of the \Halpha line on dust. Thus, the \Halpha emission we analyse here is the inherent (or, equivalently, the optically thin/absorption-corrected) emission from the gas.

\begin{figure*}
	\includegraphics[width=0.85\textwidth]{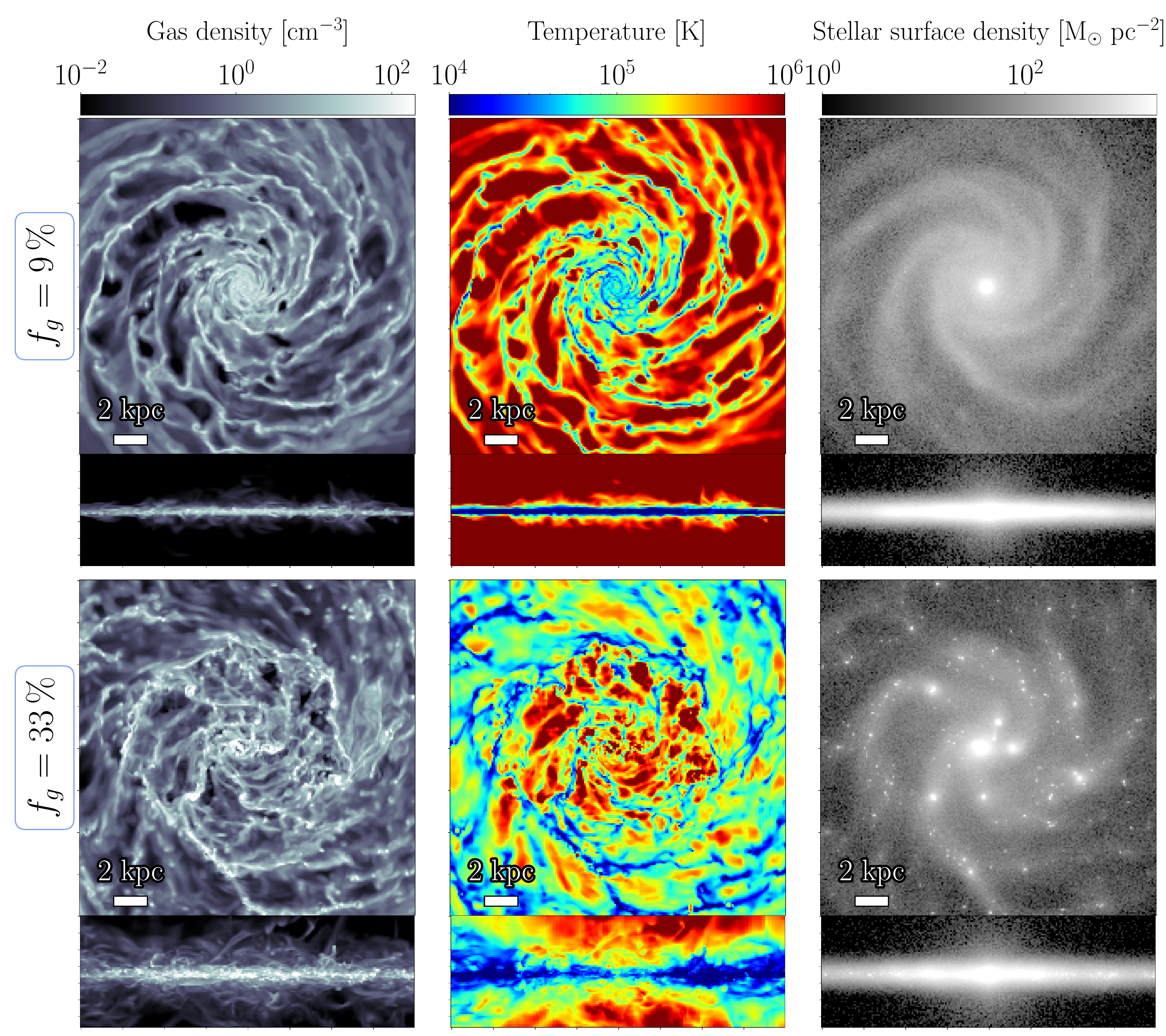}
    \caption{Projection maps of the gas and stellar properties within the simulated galaxies. Each map is projected both face- and edge-on.}
    \label{fig:density_temp_stars}
\end{figure*}

\begin{figure*}
    \includegraphics[width=0.90\textwidth]{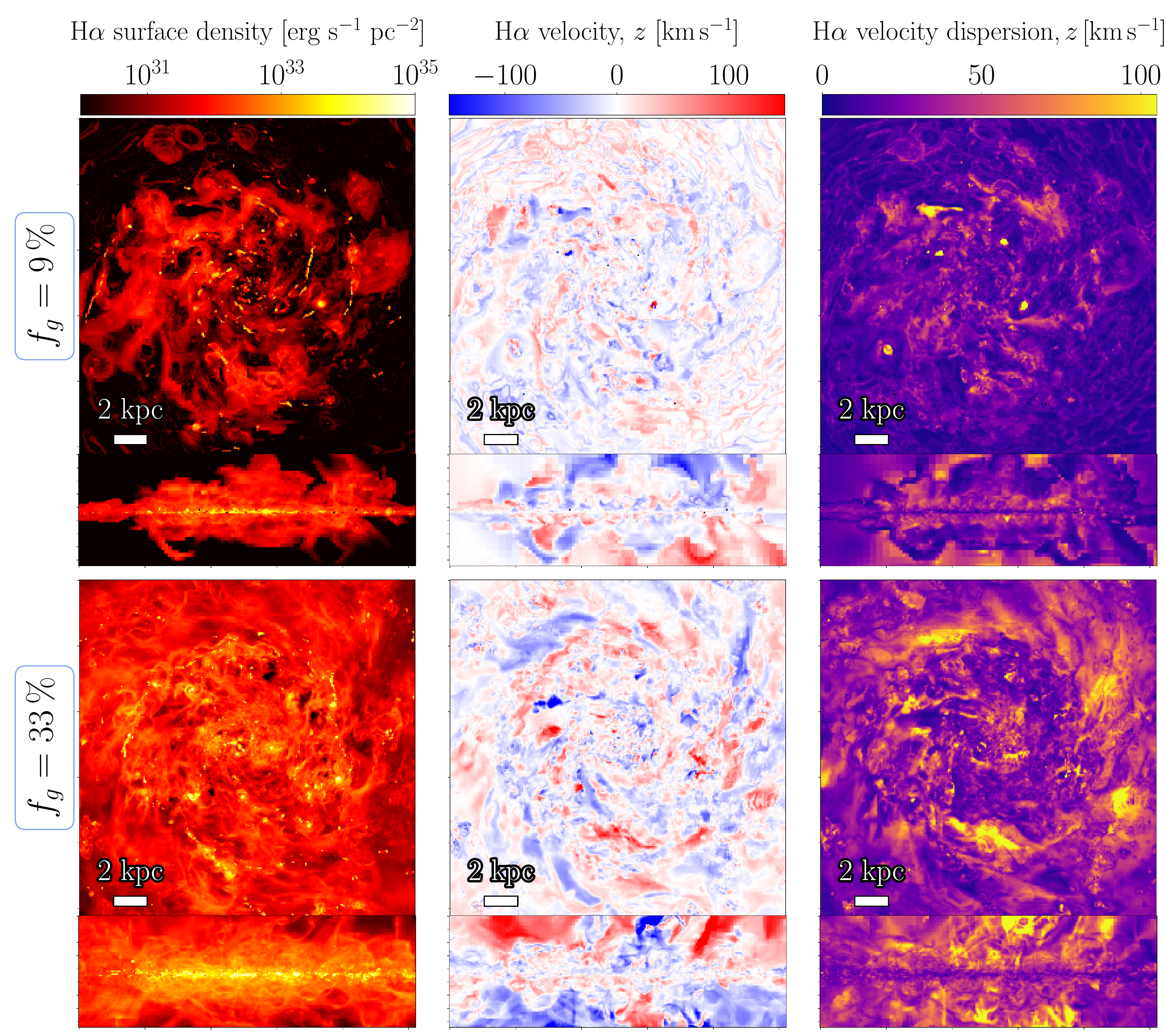}
    \caption{Projection maps of the \Halpha surface density, velocity, and velocity dispersion. The velocities are all along the $z$-axis, i.e. as if observing the galaxy face-on. The red colour in the face-on velocity projections represent gas moving towards us. The edge-on projections have their $z$-axis pointed upwards in the plot, i.e. gas above/below the disc with a positive/negative $z$-velocity are outflows. A few regions have high dispersion but low surface density ($\Sigma_{\rm H\alpha}\leq10^{30}\,{\rm erg\,s^{-1}\,pc^{-2}}$) and, thus, do not accurately represent the \Halpha kinematics.}
    \label{fig:Ha_dens_kinematics}
\end{figure*}

However, even assuming our approach is comparable to observations of dust-corrected \Halpha emission, the exact effects of dust-scattering on the \Halpha emission line is complex. In the RHD simulation by \citet[][]{Tacchella2022}{}{}, the authors performed post-processing of the \Halpha emission and showed that scattering allows 20-40\% of their \Halpha emission to escape dust-obscured regions, which increases the \Halpha contribution from dusty \HII regions and the relative \Halpha fraction between DIG and \HII regions. Furthermore, they show that scattering boosts the \Halpha signal in extraplanar DIG gas. We discuss the potential impact of not including \Halpha dust scattering on our results in Section~\ref{sec:DIGvsHII}.

In order to adequately reproduce the \Halpha signal, the spatial resolution has to be high enough to capture the smallest \HII regions. \citet[][]{Deng+23}{}{} showed how unresolved \HII regions in RHD simulations warm up the surrounding gas and lead to an unrealistic mix of cold and warm gas, resulting in a partially ionised and warm gas ($\sim$8\,000\,K) which is then overproducing \Halphaalt-bright gas, causing an over-representation of dense \HII regions to the total \Halpha signal. In Appendix~\ref{appendix:stromgren}, we carry out a post-process analysis of how well resolved, in terms of the Str{\"o}mgren radius (${\rm R_{SG}}$), photoionised gas is, and investigate the impact of resolution requirements in \HII regions. We find that unresolved, hence unrealistic, \HII regions can be filtered away by requiring ${\rm R_{SG}}$ to be resolved by at least 10 cells \citep[in line with the criterion identified by][]{Deng+23}, which we only enforce in cells with recent star formation, to avoid filtering less \Halpha bright, non-\HII regions. Cells which fit this criterion have their \emHa reduced to the mean value of resolved \HII regions in our simulations, see Appendix~\ref{appendix:stromgren} for details.

After filtering the unresolved \HII regions, we achieve a good match for the global \Halpha luminosity ($L_{\rm H\alpha}$, see Table~\ref{tab:Halpha_values}) compared to what is predicted from their star formation rate (SFR; see Table~\ref{tab:ICs}); by adopting the $L_{\rm H\alpha}$-SFR relation 
\begin{align}
    L_{\rm H\alpha}\, [{\rm erg\,s^{-1}}] = 1.26\cdot 10^{41} \,\times\, {\rm SFR\,[M_\odot \, yr^{-1}]}
\end{align}
from \citet[][]{Kennicutt98}{}{}. The predicted values for \texttt{fg10} and \texttt{fg50} are $2.5\cdot 10^{41}\ {\rm erg\,s^{-1}cm^{-3}}$ and  $5\cdot 10^{42}\ {\rm erg\,s^{-1}cm^{-3}}$, respectively.

\begin{figure*}
    \includegraphics[width=1.\textwidth]{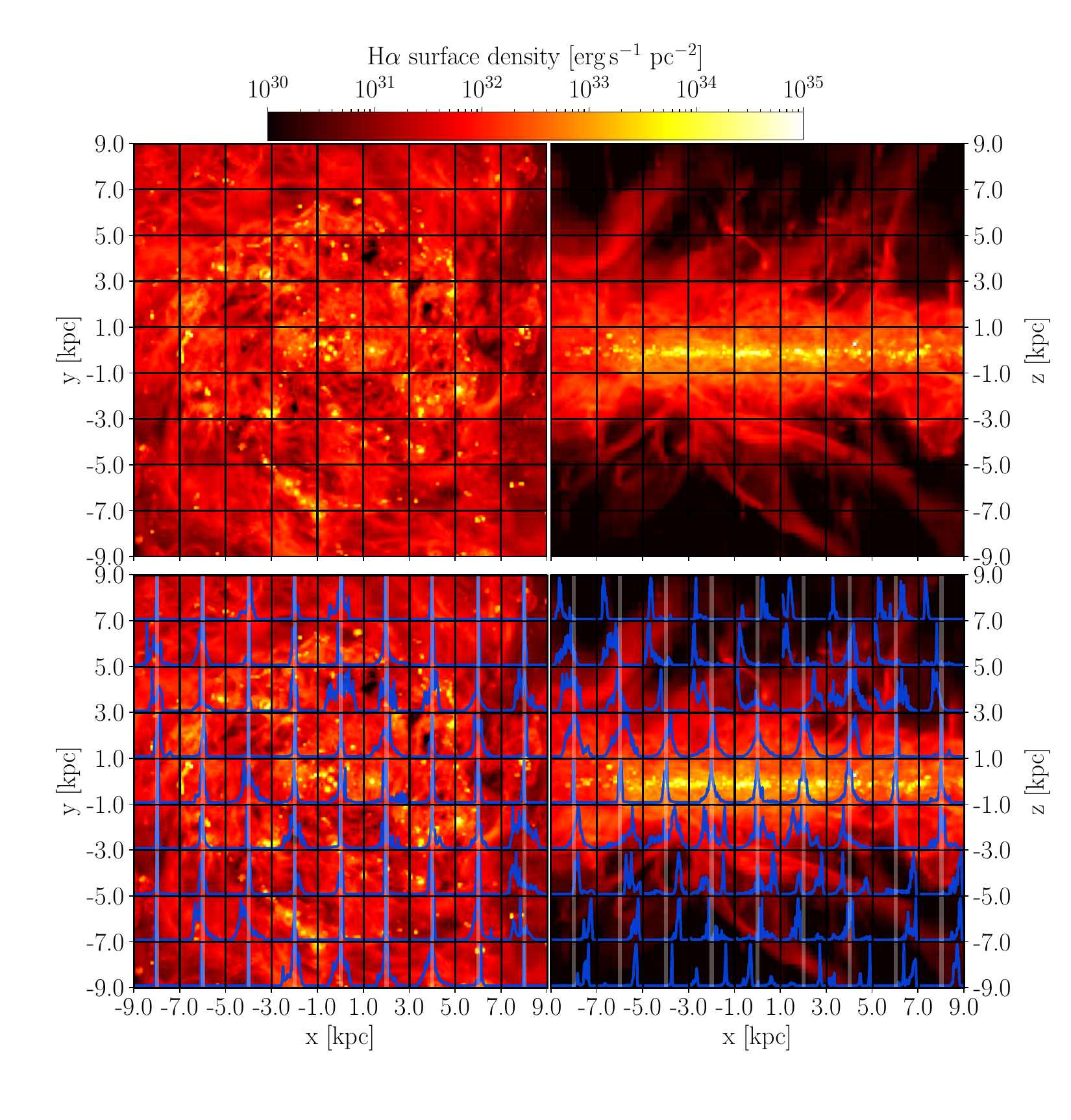}
    \vspace{-15mm}
    \caption{Projection maps, edge- and face-on, of the \Halpha emissivity for the \texttt{fg50} simulation 200\,Myr after the last refinement step (see Section~\ref{sec:calculations}). The lower maps have The blue lines in each grid patch is the normalised mock line profile, with velocity range $-200$\kmsec\ to $+200$\kmsec, of the \Halpha emission viewed along the $z$-direction. The line centre (0\,km\,s$^{-1}$) is annotated with a transparent white stripe in each grid. The edge-on projection shows clear blue/red-shifted profiles above the gas disc, which is indicative of outflows.}
    \label{fig:ha_line_map_fg30_normalised}
\end{figure*}

In order to illustrate how our \Halpha emissivity calculations perform, and to highlight the gas environment in which the \Halpha signal is the strongest, we show in Figure \ref{fig:Halpha_phase} temperature-density diagrams for each simulation, using one output 200 Myr after the last refinement (see Section~\ref{sec:calculations}). The left plots are coloured by the \Halpha emissivity and the right plots show the total volume. The left plot can be thought of as a visualisation of the weighting mask that is applied to the data when we calculate \Halpha variables. The strongest \Halpha signal coincides with the temperature- and density-range expected inside \HII regions ($T\sim10\,000$\,K, $10{\rm\, m_H\, cm^{-3}}\leq \rho\leq 1\,000\,{\rm m_H\, cm^{-3}}$). The diffuse gas ($T\sim8\,000$\,K, $\rho\lesssim 0.1\,{\rm \, m_H\, cm^{-3}}$) has notably smaller \emHa but fill a much larger volume and so it still has a significant contribution to the total \LHaalt. 

The \Halpha properties were calculated by dividing the galaxy into a square grid with 1 x 1 kpc$^2$ patches (see e.g. Figure~\ref{fig:ha_line_map_fg30_normalised} for a visualisation of a 2 x 2 kpc$^2$ grid). The velocity dispersion of the ionised gas as traced by \Halphaalt, \sigmaHaalt, was calculated (integrated) within each individual grid patch as a weighted dispersion
\begin{align} \label{eq:sigma_Ha}
    \sigma_{\rm H\alpha} = \sqrt{ \frac{\sum \epsilon_{\rm H\alpha, i}\, (v_{\rm H\alpha, i} - \bar{v}_{\rm H\alpha} )^2 }{\sum \epsilon_{\rm H\alpha, i} } },
\end{align} % {\rm H\alpha}
where $i$ denotes individual simulation cells within the patch and $\bar{v}_{\rm H\alpha}$ is the \emHaalt-weighted mean velocity within the patch. Note that we, in this work, are only considering the face-on/vertical velocity component, i.e. $z$-axis, and, thus, correcting for the rotation of the disc (and its associated beam smearing) is not relevant. As there are no additional physical or observational effects considered (except possibly spatial resolution), we are essentially calculating the intrinsic velocity dispersion.

We computed mock line profiles of the \Halpha signal within grids by binning the cell velocities $v_{\rm H\alpha, i}$, from $-200$\kmsec to $+200$\kmsec, and summing up the total \Halpha luminosity in each velocity bin. The bin size of this analysis was chosen as 5\kmsec as it resolves the average \sigmaHa in \HII regions. The choice of spectral and spatial resolution have an effect on resolving the line-width \citep[see e.g.][]{Lelli+23}{}{}, but any \sigmaHa values presented in this work use Eq. \ref{eq:sigma_Ha}, which only depend on the spatial size (beam smearing) and vary by less than a factor of 2 for a large range (see Figure 4 in Paper I). To assure that \sigmaHa is representative of observing the line-width, we compare \sigmaHa from Eq. \ref{eq:sigma_Ha} with the full-width half-maximum of each line profile. We find that on average, excluding complex profiles, the two methods are in agreement. % [If adding analysis of line-width vs velocity dispersion calculations.]
% using the \Halpha emissivity as the weights $w_i=\epsilon_{\rm H\alpha}$ and 

The \Halpha variables presented in Table~\ref{tab:Halpha_values} are time-averaged values for the entire galaxy. We calculated \sigmaHa by first calculating the local \sigmaHa within each grid patch and then computing an \Halpha weighted-mean value of these values. This method of calculating a global mean \sigmag is commonly done in observations when the galaxy is resolved, but the choice of weights varies between e.g. SFR, flux, no weights \citep[e.g.][see also the discussion in Section 2 of Paper I]{Epinat+09, Lehnert+13, Moiseev+15, Varidel+16}{}{}. The grid size adopted for the local \sigmaHa is 1\,kpc, unless otherwise stated, and we briefly explore the impact of grid size on \sigmaHa in Section~\ref{sec:DIGvsHII}.

\begin{figure*}
    \includegraphics[width=0.49\textwidth]{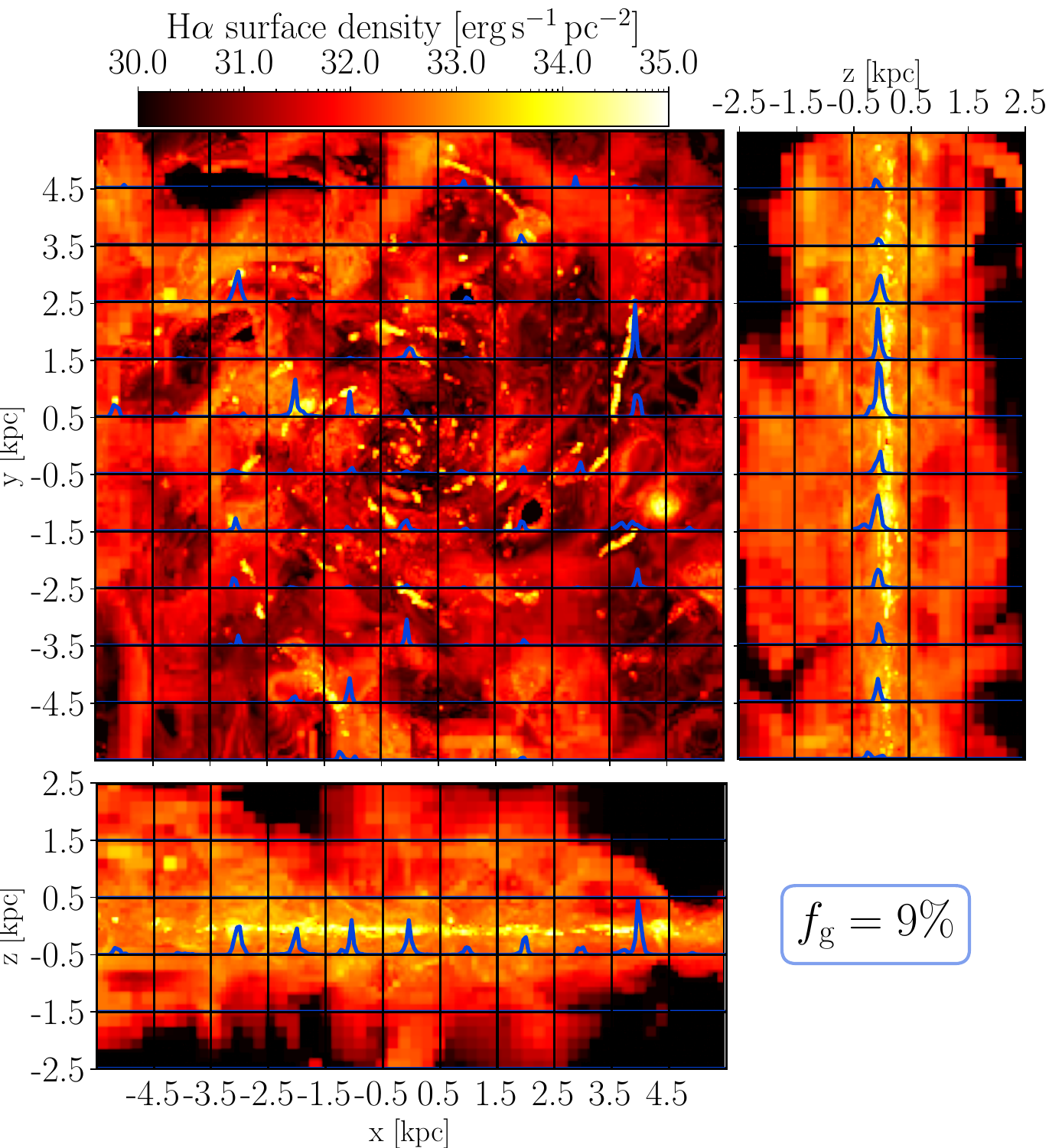}  
    \includegraphics[width=0.49\textwidth]{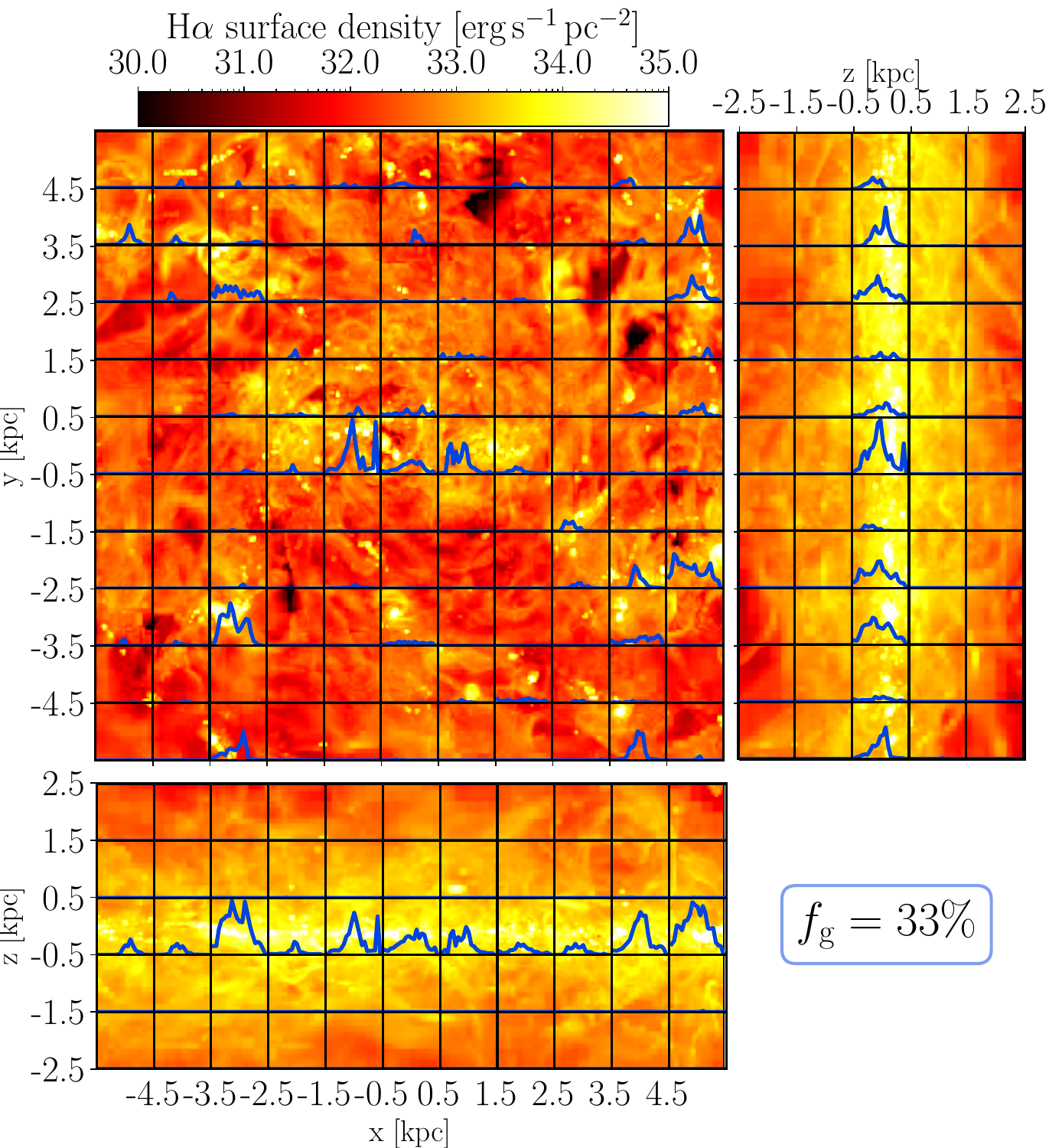}
    \caption{The \Halpha surface density maps, viewed from three angles, of the \texttt{fg10} (left) and \texttt{fg50} (right) simulations. Each map is annotated with mock \Halpha emission line profiles, with 1 x 1 kpc$^2$ grid patches, plotted as blue lines. The line profiles in all projection maps are along the vertical, $z$-direction, integrated along line-of-sight of the grid patches in the velocity range $v_{\rm z} = [-50\,{\rm km\,s^{-1}}, +50\,{\rm km\, s^{-1}}]$ with $5$\kmsec\ resolution. See Section~\ref{sec:calculations} for how these mock profiles are calculated.} 
    \label{fig:Ha_line_map_both}
\end{figure*}

\begin{figure*}
    \hspace{-15mm}\includegraphics[width=0.90\textwidth]{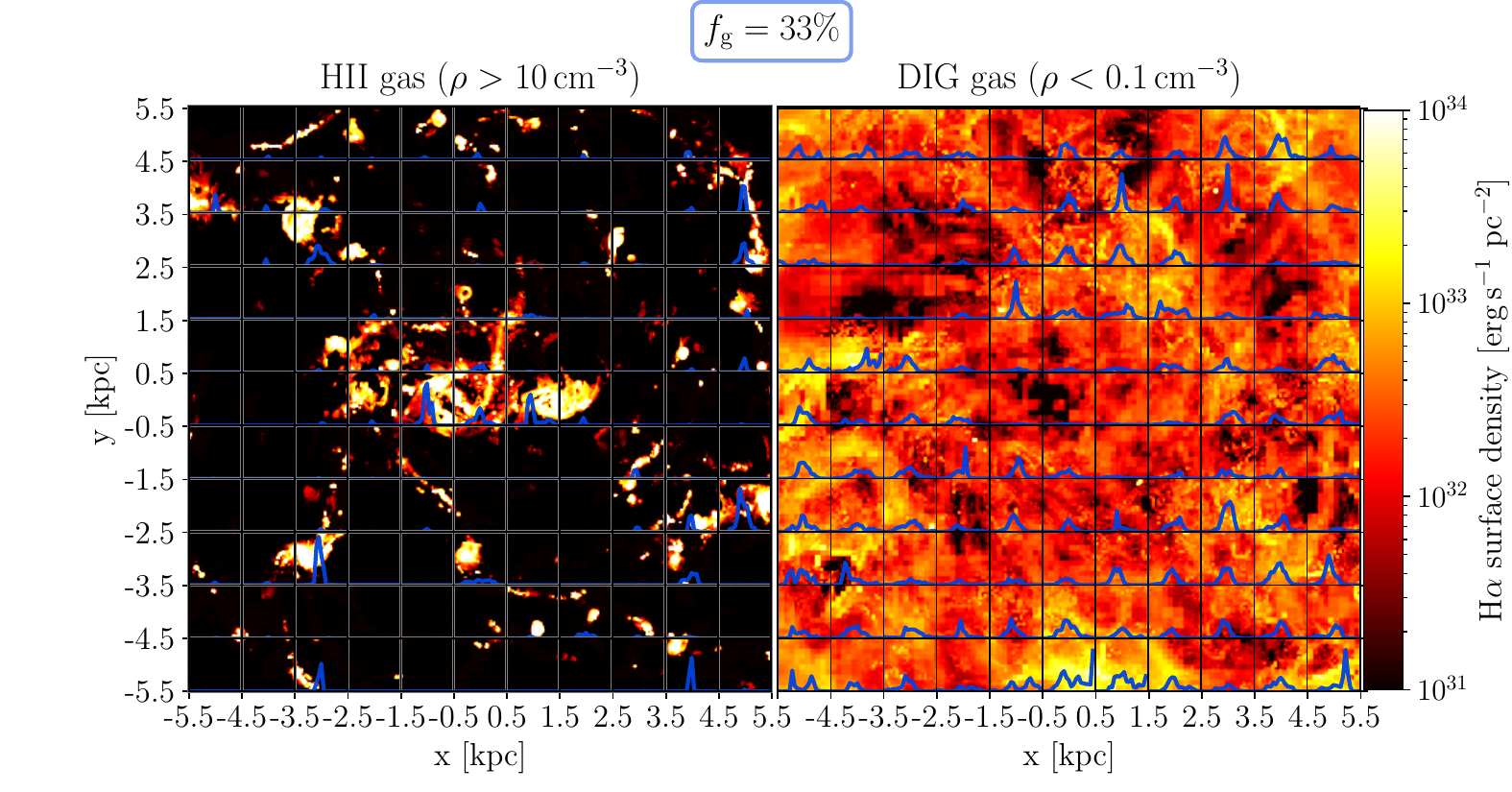}\\
    \includegraphics[width=0.85\textwidth]{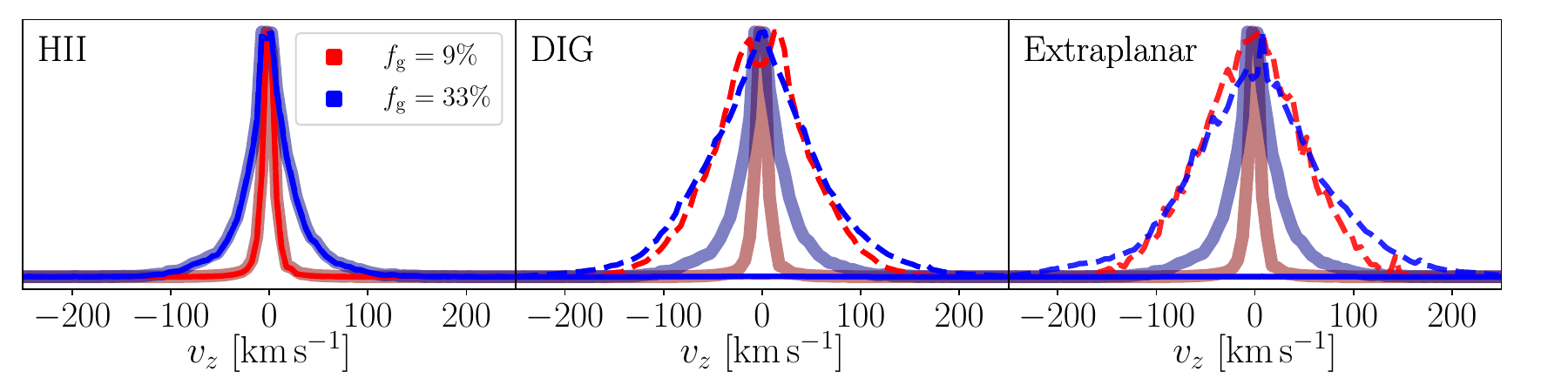}     
    \caption{{\it Top:} Projection maps of \Halpha emission in the dense (left) and the diffuse (right) gas with corresponding \Halpha mock profiles annotated on top in blue, in a 1 x 1 kpc$^2$ grid. {\it Bottom:} The \Halpha profiles integrated over the entire galaxy. The three plots split the \Halpha profile into three components: \HII regions, DIG, and extraplanar gas. The total emission of the gas-rich and gas-poor galaxy (thick transparent lines in each plot) are normalised to the same maximum. The dashed lines show the profiles normalised to the same max value as the total emission line.} % $\rho\geq0.1\,{\rm cm^{-3}}$; DIG % (similar to Figure \ref{fig:Ha_line_map_both})
    \label{fig:Ha_map_DIGvsHII} 
\end{figure*}

%%%%%%%%%%%%%%%%%%%%%%%%%%%%%%%%%%%%%%%%%%%%%%%%%%%%%%%%%%%%%%%%%%%%%%%%%%%%%%%%%%%%%%%%%%%%%%
\section{Results}
In Figure \ref{fig:density_temp_stars} we show the gas density, gas temperature, and stellar mass density maps for our simulations. A significant difference between the gas fractions is their gas and stellar structure. The lower gas fraction has its cold, dense gas (and therefore also star formation) confined within spiral arms while the higher gas fraction shows more clumpy gas and stellar structure, as found in other theoretical work on gas rich galaxies \citep[e.g.][]{Agertz+09,Bournaud2009,Renaud+21} and observed in high redshift disc galaxies \citep[e.g.][]{Elmegreen+07, Genzel+11, Zanella+19}{}{}. The maps also reveal that the higher gas fraction galaxy has a more prominent vertical structure, with a high covering fraction of warm gas ($10^4\,{\rm K}<T<10^5\,{\rm K}$) several kpc above the disc plane due to stellar feedback processes ejecting gas out from the disc in outflows, which are either e-accreted or escape into the edges of the halo \citep[see e.g.][for a review]{Veilleux+05}{}{}.

Additionally, in Figure~\ref{fig:Ha_dens_kinematics} we show the same structure of projections maps but of the surface density, velocity, and velocity dispersion, for the \Halpha emitting gas (see Section~\ref{sec:calculations} for how \Halpha properties are calculated). The kinematics are all analysed along the $z$-axis, meaning that edge-on velocity projections visualises the out- and in-flow of \Halpha gas; above the disc, positive/negative velocities (red/blue) indicates out/in-flows and vice-versa. By plotting these quantities together, it allows a visual comparison between regions with bright \Halpha emission and their corresponding kinematic features. For example, the $f_{\rm g}=33\%$ disc has a region of high \sigmaHa near the bottom centre of the disc which is slightly offset from a line of \Halpha bright \HII regions. This could be due to an older star trailing behind the \HII regions and then undergoing a supernova, which could explain the outflow seen in the velocity map. %, or photons and winds leaking from the \HII regions (as is a theory suggested for DIG to be ionised, by HII regions).
%% the rotating spiral arm (i.e.

In the coming sections, we show how warm, ionised, (\Halphaalt-bright) gas is ejected from the disc, and present resolved \Halpha mock profiles in grid patches within and outside the disc. In Section~\ref{sec:Ha_extraplanar}, we investigate in more detail the \Halpha vertical disc structure and in Section~\ref{sec:mass_loading} the amount of gas outflow.

\begin{figure}
	\includegraphics[width=0.45\textwidth]{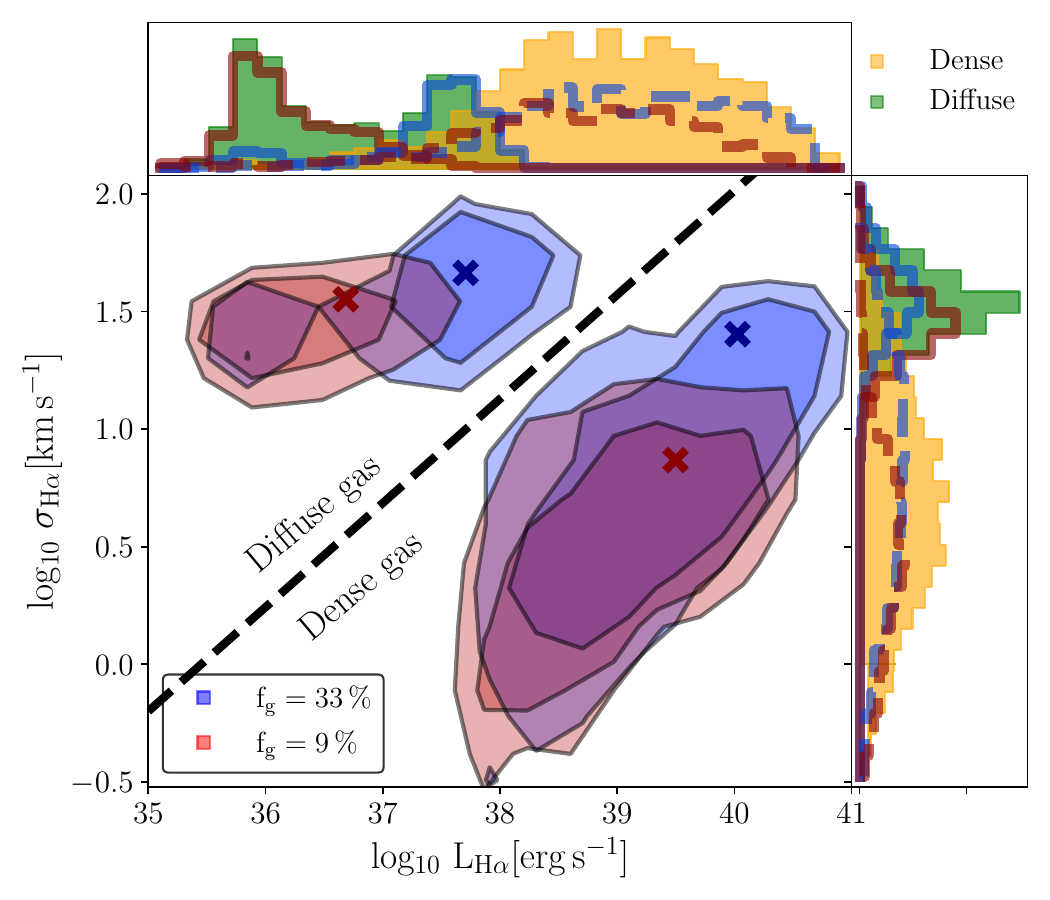}
    \caption{Contours of the relation between the \Halpha velocity dispersion and the \Halpha luminosity. The values were calculated within 1 kpc grids, as described in Section~\ref{sec:calculations}, and the contours encompass 90\,\% and 68\,\% of the data. The crosses are the \LHaalt-weighted mean. The gas is divided into diffuse ($\rho_{\rm g}\leq0.1\,{\rm cm^{-3}}$) and dense gas ($\rho_{\rm g}\geq10\,{\rm cm^{-3}}$), which creates two distinct region in this parameter space; highlighted with a dashed line. This indicates a clear separation in \Halpha luminosity and kinematics between (diffuse) DIG and (dense) \HII regions. The histograms show the distribution of the diffuse and dense gas, with line-histograms dividing it into the two simulations. }
    \label{fig:sigma_LHa_diffuse_dense}
\end{figure}

%%%%%%%%%%%%%%%%%%%%%%%%%%%%%%%%%%%%%%%%%%%%%%%%%%%%%%%%%%%%%%%%%%%%%%%%%%%%%%%%%%%%%
\subsection{The H$\alpha$ line profile}\label{sec:Ha_maps}
In this section, we present mock \Halpha line profiles (see Section\,\ref{sec:calculations} for details) for different spatial locations, inside and outside of the disc, to better understand the origins of profile shapes. In Figure \ref{fig:ha_line_map_fg30_normalised} we show the \Halpha surface density, $\Sigma_{\rm H\alpha}$, maps for the $f_{\rm g}=33\,\%$ simulation projected face- and edge-on ($z$ and $x$ projection). The maps are calculated as $\Sigma_{\rm H\alpha} = \sum L_{\rm H\alpha,i}/\Delta s^2$, with the spatial resolution $\Delta s \approx 10$\,pc. A 2\,x\,2 kpc$^2$ grid is annotated on top, showing the line-profiles in each grid, normalised to the same maximum line peak. Both projections show the line profiles in the vertical $z$ direction for all gas within that grid. In the face-on projection (left), there is no clear spatial correlation with the line-width of the profile and the \Halpha bright regions. This is partially due to the grid being too coarse to separate individual regions and upcoming maps will be more zoomed in with 1\,x\,1\,kpc$^2$ grids. The edge-on projection (right), shows that extraplanar line profiles are much broader than those inside the disc. Furthermore, extraplanar line profiles are increasingly blue- and red-shifted with increasing distance away from the disc, which is because the faster (more energetic) the outflow is, the further out it can travel. This highlights that even if the outflows are not contributing significantly to the total \LHa (as we argue below), they shape the wings of the observable profile, which we briefly quantify in Section~\ref{sec:DIGvsHII}.

The non-normalised line profiles inside of the $f_{\rm g}=9\,\%$ and $f_{\rm g}=33\,\%$ galaxies can be seen in Figure\,\ref{fig:Ha_line_map_both}. In each figure, the galaxy is projected along three different axes, but the line profiles shown for the edge-on projections are all along the $z$-axis, i.e. as if observed face-on. This method allows us to evaluate the contribution to the face-on \Halpha line profile from gas at different heights and environments within the galaxy. The face-on projections show that regions with the highest levels of $\Sigma_{\rm H\alpha}$ coincide with strong emission lines, as expected. The side-projections reveal that the total line profile is completely dominated by emission from warm gas residing in the midplane of the disc, even for the gas rich galaxy which has roughly 20 times stronger gas outflow than the low gas fraction (see Section~\ref{sec:mass_loading}). We further quantify the \Halpha emission and line-width variation with height in Section~\ref{sec:Ha_extraplanar}.

\subsection{Decomposition of the \Halpha profile}\label{sec:turbulence}
In this section we split the \Halpha signal into DIG ($\rho<0.1\,{\rm cm^{-3}}$), \HII regions ($\rho>10\,{\rm cm^{-3}}$), and extraplanar ($|z|\geq1.0$) components to evaluate their contribution to the total \Halpha profile. In Figure~\ref{fig:Ha_map_DIGvsHII}, we show projection maps of the \HII and DIG phases for $f_{\rm g}=33\%$ with their corresponding \Halpha line profiles annotated in 1x1\,kpc$^2$ patches (same as previous figures). Below the projections, we show plots of each component of the \Halpha profile, summed/integrated over the entire galaxy, for both the low and high gas fraction galaxy. These can then be compared to the total, combined, \Halpha profiles of both galaxies (the thicker, transparent lines) which are normalised to the same height. Additionally, we present the same profiles but normalised to the maxima of the total profiles (dashed lines), to compare their relative shape.

As to be expected for \HII regions, the \Halpha bright and dense gas is confined within clumps and spiral arms, while the diffuse gas is more homogeneously spread throughout the galaxy. This can also be seen from the annotated line profiles, which are strongest near the clumps for \HII regions and roughly equal in strength for the DIG.

% \textbf{State the relative amount of DIG/HII inside and outside of the disc. Allude here to how much of the gas is in the wings and refer to [small discussion in] Section~\ref{sec:Ha_outflows} for the amount of gas above 200kmsec, as it is related to outflows? Or maybe just mention here...}

For the \Halpha line profiles integrated over the entire galaxy, the \HII profiles are very narrow and follow the total \Halpha profiles very closely (which makes them difficult to distinguish), meaning that \HII regions constitute the majority of \Halpha emission even at velocities $\gtrsim50$\kmsec. Comparatively, the DIG profile is rather weak and contributes only roughly $f_{\rm DIG}\leq10\%$ of the total \LHaalt. However, from the normalised DIG line (dashed lines) it is clear that the \Halpha emission line from the DIG is much broader. In Section~\ref{sec:DIGvsHII}, we quantify the \Halpha line profiles of DIG and \HII regions in terms of their \LHa and velocity dispersion \sigmaHaalt, as a proxy for line-width, and show how these properties evolve with gas fraction.

Furthermore, the extraplanar gas provides at most a few percent of the total \Halpha emission, but is broader, as can be seen from the normalised line, and thus has notably higher \sigmaHaalt. Additionally, the extraplanar gas makes up $\sim45\%$ of the total (i.e. both inside and outside the disc) \Halpha profile extended wing ($\geq200$\kmsec). We further quantify the vertical structure of \Halpha emission and turbulence in Section~\ref{sec:Ha_extraplanar}. Additionally, $\sim$25\%\,/\,55\% of the DIG is outside 1\,kpc of the disc in the low/high gas fraction galaxy and makes up essentially all of the extraplanar \Halpha emission, which explains the similarity between the DIG and extraplanar profile. In the high gas fraction galaxy, most of the \LHa at high-velocities $v_z\sim100$\kmsec comes from high-density gas $\rho\sim 10\,{\rm cm^{-3}}$ in the disc, which is likely feedback accelerated gas that has yet to be ejected. Although, the DIG constitutes 80\% of \LHa above 200\kmsec within the disc, which means it makes up majority of the \Halpha profile's extended wing. Thus, extraplanar gas is mainly observed in the wings and at larger heights the \Halpha lines become blue/red shifted (see Figure~\ref{fig:ha_line_map_fg30_normalised}), an indication of in/out-flow. The amount of outflow in our galaxies is discussed more in Section~\ref{sec:mass_loading}.

% Essentially all of the dense/HII gas ($\rho>10\,{\rm cm^{-3}}$) resides within the disc, while the low/high gas fraction galaxy has 23\%/56\% of the DIG \Halpha luminosity outside of the disc. 

% Furthermore, in the high gas fraction galaxy, the extraplanar \Halpha profile is stronger, which is expected from a thicker gas disc or more massive gas outflows; although, it is also more narrow due to the disc thickness. 

% Each figure also contain the total \Halpha line profile of the galaxy as a thick black line, as well as divided into specific components: the extraplanar gas ($|z|>1.0$ kpc), \HII regions, and the DIG. The dashed lines are the profiles of the component corresponding to that colour, but normalised to the peak value of the total \Halpha profile. 

%%%%%%%%%%%%%%%%%%%%%%%%%%%%%%%%%%%%%%%%%%%%%%%%%%%%%%%%%%%%%%%%%%%%%%%%%%%%%%%%%%%%%%%%%%%%%%
\section{Discussion}
In this Section, we investigate in more detail the variation in \sigmaHa in regions with diverse temperatures and densities. Next, we quantify the amount of gas being ejected away from the disc, using the mass loading factor, and explore the contribution of extraplanar gas and outflows to the total \Halpha emission.

%%%%%%%%%%%%%%%%%%%%%%%%%%%%%%%%%%%%%%%%%%%%%%%%%%%%%%%%%%%%%%%%%%%%%%%%%%%%%%%%%

\begin{figure*}
    \includegraphics[width=0.70\textwidth]{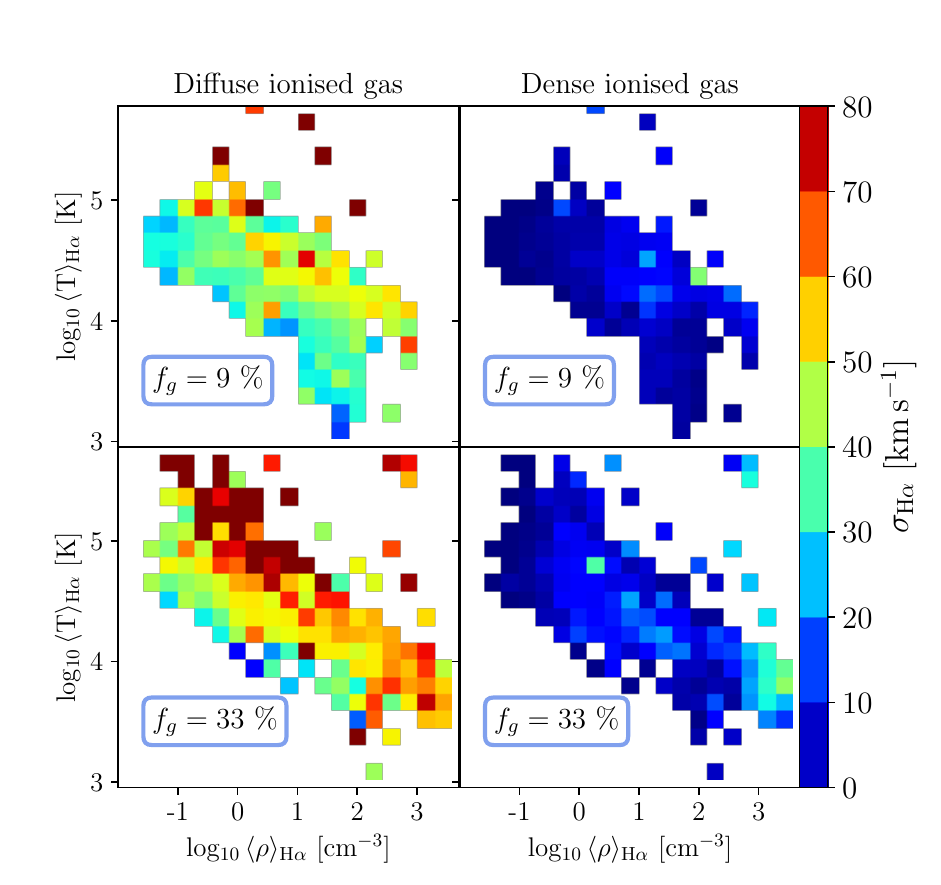}
    \caption{Temperature-density phase diagrams coloured by the \Halpha velocity dispersion for our simulations. The temperature $\langle T \rangle_{\rm H\alpha}$ and density $\langle \rho \rangle_{\rm H\alpha}$ presented in this figure are \emHaalt-weighted averages of all the simulation cells within each grid. The kinematics (but not the temperature or density) were calculated by separating the \Halpha gas into diffuse gas (DIG) and dense gas (\HII regions). }
    \label{fig:phase_diagram_sigma}
\end{figure*}

%%%%%%%%%%%%%%%%%%%%%%%%%%%%%%%%%%%%%%%%%%%%%%%%%%%%%%%%%%%%%%%%%%%%%%%%%%%%%%%%%%%%%%%%%%%%%%
\subsection{The ${\rm H\alpha}$ turbulence in DIG and H{\small II} regions}\label{sec:DIGvsHII}
Observations of the ionised gas with the \Halpha line have shown that disc galaxies are supersonically turbulent systems, with velocity dispersions $\sigma_{\rm H\alpha}\approx 10-100$\kmsec (see discussion and references in Section~\ref{sec:intro}). However, there is a large spread in \sigmaHa between observations and it is not clear how the turbulence varies for gas in different physical states and local environments. We investigate the multi-phase nature of \sigmaHa by dividing the gas into dense ($\rho_{\rm g}\geq10\,{\rm cm^{-3}}$) and diffuse gas ($\rho_{\rm g}<0.1\,{\rm cm^{-3}}$), to represent \HII regions and the DIG, respectively\footnote{We find that the exact choice of limits does not have a significant impact on the \LHa or \sigmaHa values we calculate. Additionally, using the electron density $n_e$ instead of gas density does not change our results.}.

In Figure~\ref{fig:sigma_LHa_diffuse_dense}, we plot \sigmaHa as a function of \LHa in each grid patch of our two simulations. The plot is divided into two distinct regions based on their density, which is highlighted with a dashed line. Additionally, the figure shows histograms of the \LHa and \sigmaHa distribution for the diffuse and dense gas. The diffuse gas is able to reach velocity dispersions as high as \sigmaHa$\lesssim 80$\kmsec within individual patches and has a mean value of \sigmaHa$\sim50$\kmsec. The denser, more luminous, gas is less turbulent with $\sigma_{\rm H\alpha}\lesssim 40$\kmsec and an average value of $\sigma_{\rm H\alpha}\sim 29$\kmsec. There is also an evolution with increased gas fraction, with the gas rich galaxy ($f_{\rm g}=33\%$) being more luminous and turbulent than the gas poor case ($f_{\rm g}=9\%$) in \emph{both} the diffuse and dense ionised gas. This increase in \sigmaHa with gas fraction is likely driven by the higher SFR (or $\Sigma_{\rm SFR}$) in the high gas fraction galaxy, as we found the same evolution in Paper I, both in individual patches and averaged over the entire galaxy.

As this figure illustrates, the \Halpha emission originates from, at least, two different gas phases with distinct \LHa and \sigmaHaalt. Separating the gas into DIG and \HII is important for disentangling the \Halpha emission line and deriving accurate physical properties of the individual gas phases \citep[see e.g.][for observational examples of DIG studies where this is done]{DellaBruna+22, Micheva+22}{}{}. In Section~\ref{sec:DIGvsHII}, we go into more detail about how \sigmaHaalt, for the diffuse and dense ionised gas, varies in different density-temperature environments.

%%%%%%%%%%%%%%%%%%%%%%%%%%%%%%
%%%%%%%%%%%%%%%%%%%%%%%%%%%%%%%%
%%%%%%%%%%%%%%%%%%%%%%%%%%%

The contribution of \Halpha emission from different gas phases can be understood from the left plots of Figure~\ref{fig:Halpha_phase}, which indicate that the majority of the \Halpha emission comes from a temperature range around $T\sim 10^3 - 10^5$\,K, and a wide range of gas densities $\rho\sim10^{-2} - 10^3\,{\rm cm^{-3}}$. In Section~\ref{sec:turbulence} we divided the data into diffuse and dense gas to represent the DIG and \HII regions, respectively, and highlighted how the dense ionised gas is less turbulent than the diffuse gas. While the dense \HII regions are overall more luminous they also suspend a small volume of the ISM and might thus not represent the overall ionised gas kinematics. Here, we explore how \sigmaHaalt varies across these wide temperature and density ranges.

In order to investigate the physical state and turbulence levels of \Halpha emitting gas, we show the gas velocity dispersions, binned in temperature-density phase space, in Figure~\ref{fig:phase_diagram_sigma}. The temperatures and densities presented in this figure is the \emHaalt-weighted averaged of all the simulations cells in each 1x1 kpc$^2$ grid patch. Notably, this means that the (average) gas density is shifted to higher densities compared to the phase diagram of individual simulation cells in Figure~\ref{fig:Halpha_phase}. We calculate \sigmaHa for the diffuse and dense gas separately, but in the same environment; i.e. the temperatures and densities represent mean values of \emph{all} the gas, but \sigmaHa is calculated only for the diffuse or dense gas ($\rho\leq0.1\,{\rm cm^{-3}}$ and $\rho\geq10\,{\rm cm^{-3}}$, respectively).

The dense gas exhibits turbulence levels of $\sigma_{\rm H\alpha}\sim 10-20$\kmsec, except for in very high density patches in the high gas fraction galaxy, in which it can reach up to $\sigma_{\rm H\alpha}\lesssim 40$\kmsec. These high-density patches likely contain \Halpha luminous \HII regions, which boost \sigmaHaalt, and are confined in small regions. On the other hand, the diffuse gas is ubiquitous throughout (and outside) the disc and is highly turbulent even in the dense environments, i.e. around the \HII regions. The diffuse gas in the high gas fraction galaxy has $\sigma_{\rm H\alpha}\sim 50-60$\kmsec in most environments and, in the more gas rich galaxy, can reach $\sigma_{\rm H\alpha}\lesssim 80$\kmsec in patches with hotter, and more tenuous, gas. Furthermore, the DIG has been observed to be warmer ($T\sim8\,000$\,K) than \HII regions ($T\sim10\,000$\,K), which can be seen in this figure. Notably, \sigmaHa is slightly larger at the DIG temperatures and lower densities. This highlights the distinct kinematic offset between these two gas phases.

The spatial proximity between the DIG and \HII region gas phases, as is seen from observations \citep[e.g.][]{Zurita+02}{}{}, is beyond the investigations of this project, but we do note that the strongest \Halpha signals from the diffuse gas are close to the star forming regions. Additionally, the DIG turbulence near the densities of star forming regions ($\rho\sim100-1000{\rm \, cm^{-3}}$) seem to show slightly higher \sigmaHa than the DIG mean (see Figure~\ref{fig:phase_diagram_sigma}). Furthermore, we check the effects of varying the patch size from which we calculate each parameter. Notably, when using larger patch sizes \sigmaHa becomes more skewed towards the lower value associated with \HII regions as these dense regions dominate the \Halpha luminosity. If the contribution of \LHa from DIG regions was higher in our galaxies, \sigmaHa might instead increase at lower spatial resolutions.

In Figure~\ref{fig:sigma_LHa_diffuse_dense}, there is a clear correlation between \LHa and \sigmaHaalt. These are the most luminous \HII regions, which reach upwards of \sigmaHa$\approx40$\kmsec, and are regions of recent massive star formation. This connection between SFR and \sigmag is something we explored in detail in Paper I. The driver of global turbulence within galactic disc is still under discussion \citep[see e.g.][and references therein]{Krumholz+18, Ejdetjarn+22, Ginzburg+22}{}{}, but it is clear from this figure that the turbulence traced with \Halpha can be strongly tied to these dense \HII regions. This implies observations of disc galaxies similar in mass and age to those we present here, with \sigmaHa$\gtrsim40$\kmsec\, require a greater \LHa contribution from the DIG. Indeed, a similar cut-off has been found for \sigmaHa in DIG and \HII regions in large-scale surveys \citep[][]{Law+22}{}{}. As \HII regions are shrouded within molecular clouds, these dense gas phases likely feature similar gas kinematics. Indeed, the offset in \sigmaHa between DIG and \HII regions in our simulations is reminiscent of the offsets in turbulence levels reported by \citet[][]{Girard+21} for dense molecular gas (as traced by CO) and \Halpha traced ionised gas. We leave the investigation of the relative levels of turbulence between ionised and molecular gas phases to future work (Ejdetj{\"a}rn et al., in prep).

As discussed in Section~\ref{sec:calculations}, our approach to calculate \Halpha emission assumes a perfect dust-correction. Accounting for dust absorption in \HII regions in our simulations would reduce the amount of \Halpha emission from \HII regions and favour the contribution from the DIG, resulting in a boost in the relative DIG/\HII region contribution. This, in turn, would shift total \sigmaHa towards higher values (i.e. closer to \sigmaHa of the DIG). \citet[][]{Tacchella2022}{}{} find a 3-5 times higher DIG contribution $f_{\rm DIG}$ than our work, which could possibly be explained by their post-process method of propagating the \Halpha emission through the ISM, which takes into account absorption and scattering of the line. 

% As we are not using post-processing to trace emission line propagation, the \Halpha line is unaffected by dust absorption and scattering. Extraplanar \Halpha emission from dust scattering has been found by RT models to contribute $\sim5-20\%$ \citep[][]{ WoodReynolds99, Barnes+15}{}{} of the total luminosity. This was also numerically shown by \citep[][]{Tacchella2022}{}{}, who found that dust scattering contributed 10-30\% of their \Halpha emission above the disc. Thus, $\Sigma_{\rm H\alpha}$

\begin{figure}
  \centering 
    \hspace{-6mm}\includegraphics[width=0.45\textwidth]{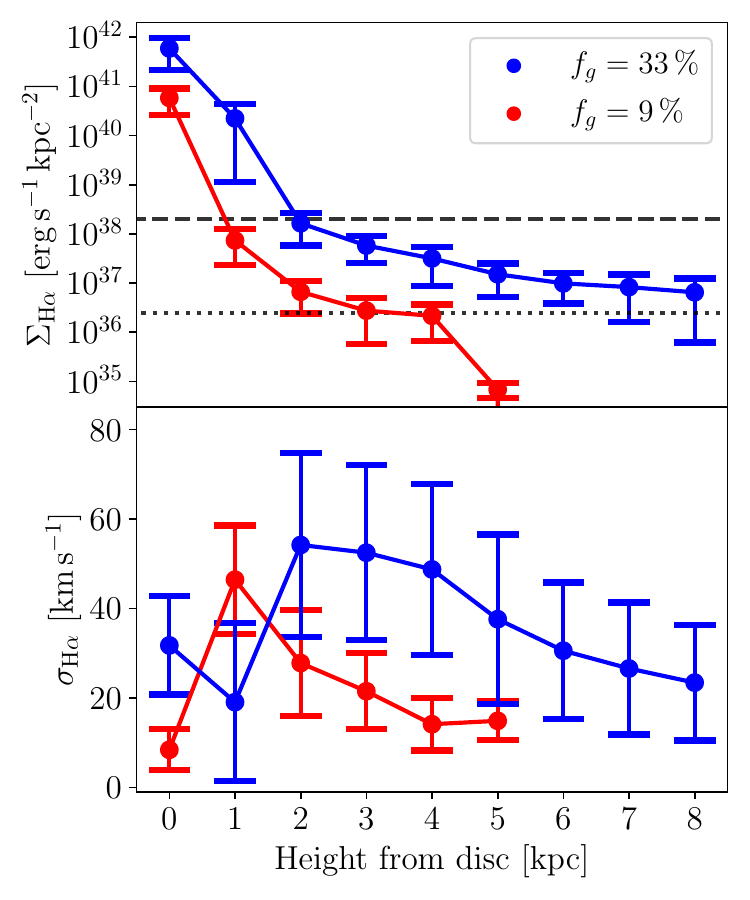}
    \caption{The \Halpha luminosity and velocity dispersion as a function of the height from the disc. The values are time-averaged from several simulation outputs with a 1$\sigma$ errorbar. A sensitivity limit, based on the capabilities of MUSE and JWST within the local/high-redshift universe (comparable to the low/high gas fraction galaxy), is plotted as a dotted/dashed grey line (see the details of how these are determined in Section~\ref{sec:Ha_extraplanar}). }
    \label{fig:Halpha_height}
\end{figure}

%%%%%%%%%%%%%%%%%%%%%%%%%%%%%%%%%%%%%%%%%%%%%%%%%%%%%%%%%%%%%%%%%%%%%%%%%%%%%%%%%%%%%%%%%%%%%%
\subsection{Is the extraplanar H$\alpha$ gas observable?}\label{sec:Ha_extraplanar}
% \subsection{What is the contribution of extraplanar gas to the H$\alpha$ signal?}\label{sec:Ha_outflows}

% Sardaneta+24: sensitivity ~ , observe down to ~ 
% Levy+19: sensitivity ~
% Ho+16: sensitvity ~ 
% In galaxies lacking spectacular large-scale winds, narrowband Hα imaging in nearby late-type edge-on galaxies reveals that eDIG is very common, with more than half of the galaxies showing extraplanar diffuse emissions and sometimes filamentary structures (e.g. Rossa & Dettmar 2000, 2003a,b; Miller & Veilleux 2003a; Rossa et al. 2004). The average distances of the extended emission above the galactic midplane can range from 1–2 to 4 kpc or more.
% MIller & Veilleux - The H scale height of the eDIG derived from a two-exponential fit to the vertical emission profile ranges from 0.4 to 17.9 kpc, with an average of 4.3 kpc.

In Section\,\ref{sec:turbulence} we showed that extraplanar gas only has a minor contribution to the total \Halpha line profile. Here, we further quantify this by calculating the average \LHa surface density ($\Sigma_{\rm H\alpha}$) and \sigmaHa at different height bins, presented in Figure~\ref{fig:Halpha_height}. To filter-out the contribution of non-\Halpha gas, we used the threshold $\epsilon_{\rm H\alpha} \geq 10^{-30}\,{\rm erg\,s^{-1}\,cm^{-3}}$, which we tested and found equivalent to performing a temperature cut $(6\cdot10^3\,{\rm K}\leq T\leq3\cdot 10^4$\,K), characteristic of \Halpha emitting gas. The values were calculated as the sum of \LHa within individual 1x1\,kpc$^2$ patches at each height and at different time outputs, and each data point on the plots is the mean values of all these patches. This approach best captures the spatial and temporal variation of $\Sigma_{\rm H\alpha}$ in individual patches (filaments) above the disc, which is represented by the error bars. The black dotted/dashed lines are theoretical limits for observations with the MUSE and JWST instruments, explained in detail in the coming paragraphs.

% but is also more directly comparable to observations than a homogeneous slab [gas outflows might only be observed as a few bright patches]

The figure shows an initial sharp decline in the \Halpha brightness for both simulations, which becomes more gradual above 2\,kpc. There is also is an initial increase in \sigmaHa at 1-2\,kpc from the disc, which can be attributed to the mix of in- and outflows at different speeds creating an extraplanar environment with a wide range of velocities. In particular, $f_{\rm g}=9\,\%$ has a sharp increase in \sigmaHa just above the disc $\sim$1 kpc while $f_{\rm g}=33\,\%$ is at 2 kpc, which is due to the lower gas fraction disc being thinner and having weaker feedback, meaning in- and outflows are closer to the disc. Further out, the dispersion decreases due to less mixing of out- and inflow.
%mostly outflows reaching far away from the disc and then slowing down, cooling, and. 

In order to assess whether the modelled extraplanar gas is observable, we consider observations of two edge-on disc galaxies at redshift $z\sim0$ and $z\sim 1.5-2.5$, appropriate for our $f_{\rm g}=9\,\%$ and $f_{\rm g}=33\,\%$ galaxy, respectively (see Section~\ref{sec:ics}). Starting with the low redshift/gas-fraction case, an efficient instrument for observing extended nebular gas is the MUSE integral field spectrograph on VLT, offering a field of view of 60”\,x\,60” \citep[][]{Bacon+10}, suitable to probe \Halpha emission in local galaxies. We perform calculations for our model galaxy assuming a redshift of $z=0.025$ for which the MUSE field-of-view corresponds to 30\,x\,30 kpc, suitable for our model galaxy and typical of the bright parts of local disc galaxies. We consider dark conditions, a total integration time of 6 hour and find that a 1\,kpc$^2$ patch can be detected at 5$\sigma$ for \Halpha surface brightness $\Sigma_{\rm H\alpha} = 5\cdot 10^{-19}\,{\rm erg\,s^{-1}\,cm^{-2}\,arcsec^{-2}}$ ($2.5\cdot 10^{36}\,{\rm erg\,s^{-1}\,kpc^{-2}}$). A 1\,kpc high and 30\,kpc wide slab is detectable at 5$\sigma$ for $\Sigma_{\rm H\alpha} = 6.6\cdot 10^{35}\,{\rm erg\,s^{-1}\,kpc^{-2}}$. This means that the extraplanar gas in the $f_{\rm g}=9\,\%$ galaxy can be seen up to heights of 4\,kpc, and even be resolved in kpc$^2$-sized patches where the emission is bright.

Observation of edge-on local disc galaxies in \Halpha have found mean scale heights of 1-2 kpc \citep[][]{Ho+16, Bizyaev+17, Levy+19, Sardaneta+24}{}{}, with individual galaxies reported as high as $> 10$\,kpc \citep[e.g.][]{MillerVeilleux03}{}{}. These observations have flux limits in the range $\sim 10^{-16}-10^{-18}\,{\rm erg\,s^{-1}\,cm^{-2}\,arcsec^{-2}}$. At those sensitivities, our low gas fraction/local analogue galaxy could be observed 'only' up to 1-2 kpc, compared to $\sim4$\,kpc based on our theoretical MUSE limits, which are 10-100 times lower. Additionally, these observations are not perfectly edge-on ($\gtrsim80^\circ$), and the disc might appear thicker compared to our simulations.
% Furthermore, with an instrument 10-100 times more sensitive (e.g. MUSE), local discs might be observable in \Halpha up towards 4 kpc.

Determining the line width requires higher signal-to-noise (S/N$\sim20$ assumed) and can be done down to $\Sigma_{\rm H\alpha}\sim 10^{37}$, which enables the brightest patches to be characterised up to $h=2$\,kpc. For an entire slab, the average line width (as in the lower panel in Figure \ref{fig:Halpha_height}) can be determined to $h=4$\,kpc. Hence characterizing the extraplanar gas distribution and width is entirely feasible in the local universe with MUSE and would allow our simulation predictions to be confronted with observations. At the assumed redshift, and in good seeing conditions, the disc and extraplanar ionised gas distribution and line width could be studied at 0.3\,kpc resolution up to heights of $\sim$1\,kpc.

Observing extraplanar disc \Halpha emission at $z=1.5-2.5$ is more challenging (despite the \emph{more luminous} $f_{\rm g}=33\,\%$ model being characteristic for the disc population at these redshifts) due to the cosmological surface brightness dimming and cosmological scale as such (where 1 arcsecond corresponds to $\sim$8\,kpc) requiring high spatial resolution. Could JWST/NIRSPEC/IFU do the job? At these redshifts its 3” x 3” field correspond to 25 x 25\,kpc$^2$, suitable for the task. With NIRSPEC/IFU and G140H/F100LP in 60\,ks integration, the limit to obtain ${\rm S/N}=5$ for a 1\,kpc$^2$ patch is $10^{39}\,{\rm erg\,s^{-1}\,kpc^{-2}}$ at $z=1.4$, which would allow sampling the disc and extraplanar gas distribution up to $h=1$\,kpc, but for the latter, only derive the line width for the brightest patches. For a 1\,kpc thick (and 3” wide) slab on both sides a surface brightness of $2\cdot10^{38}\,{\rm erg\,s^{-1}\,kpc^{-2}}$ would be detectable at $\sim 5\sigma$. This means that extraplanar \Halpha would be detectable to $h=2$\,kpc, but its line width would not be well determined. Hence, the increase in \sigmaHa predicted for $h\geq2$\,kpc will be beyond observational reach except perhaps for galaxies with higher gas fractions.

%In view of this and the fact that these are after all model \emph{predictions}, a wiser approach may be to survey a small sample of $5-6$ discs at $z\sim1.5$ with shorter exposure times ($10-12$ ks) which would fit within a small program allow probing the surface brightness and \sigmaHa distribution up to $h=1$\,kpc, and for a slab at $h=1.5$\,kpc, for a comparison with the model, and depending on the results design deeper follow-up observations.

Marginally better surface brightness limits could be obtained with NIRSPEC/MSA by forming slitlets of $\sim$7 shutter length and placing them $\sim$2\,kpc above/below the plane which could allow constraints on the line width, but would have the drawback of not sampling the surface brightness distribution vs height. On the longer time scale, ELT/MOSAIC in its mini IFU mode \citep[][]{Hammer+21}{}{} could push the detection of slabs an more than a factor of 10 fainter than NIRSPEC/IFU, but will have limited spatial resolution ($\sim$0.4” or $\sim$3\,kpc), making it more suited to study extraplanar gas at large heights. ELT/HARMONI would be suitable for studying extraplanar gas at low heights ($<3$\,kpc). We also looked into the capabilities of VLT/ERIS, but found that it would provide lower sensitivity in the same integration time. 
\subsection{The rate of gas outflow }\label{sec:mass_loading}
In this section, we investigate the presence of galactic winds and magnitude of outflows in our simulations. First, we highlight here the low fractions of galactic winds, i.e. gas outflow that escapes the galaxy halo, in our galaxies. We do this by calculating the fraction of \LHa and mass of gas exceeding the escape velocity of a MW-mass galaxy, $v_{\rm esc,\,MW}=\sqrt{2GM/r} \approx 500$\kmsec (within the disc). We find that the fraction of DIG with $v_z$ larger than half of $v_{\rm esc}$ is $\lesssim 1$\% of the disc DIG \LHaalt and $\lesssim 1$\% of the diffuse ($\rho\leq0.1\,{\rm cm^{-3}}$) gas mass. Thus, large-scale winds are not present in either of our simulations, but fountain outflows can temporarily enrich the halo with new gas. Additional acceleration, from outward pressure forces, have been suggested to alleviate the escape velocity condition to form (slowly expanding) large-scale winds, but the source of this acceleration has primarily been attributed to cosmic rays \citep[e.g.][]{Girichidis+18}{}{}. % However, exceeding the midplane escape speed is not a strict condition for gas to escape the halo, as additional acceleration at larger heights can alleviate this, e.g. from hydrodynamic, radiative \citep[][]{Thompson+05}{}{}, or cosmic ray \citep[e.g.][]{Girichidis+18}{}{} pressure forces. %Furthermore, gas with $v_z<250$\kmsec can still reach the outer part of the halo.

% The escape velocity of a galaxy with the mass of the MW is around 500\kmsec. However, it is possible to escape despite being lower than this. Even if being generous with this limit, we find that less than 1\% of gas has a velocity $v_{\rm z}\geq 250$\kmsec. Gas is still able to escape to great distances, but do not "escape" the potential well completely.
%We note that outflows exceeding the escape velocity at a particular distance and time are neither a sufficient nor necessary condition for the eventual ejection of gas outside of the DM halo. Pressure forces in the outer halo can be larger than the inward gravitational pull, effectively accelerating the gas outwards, thus alleviating the need to reach the local escape velocity (e.g. Girichidis et al. 2018).

\begin{figure}
  \centering 
    \hspace{-6mm}\includegraphics[width=0.45\textwidth]{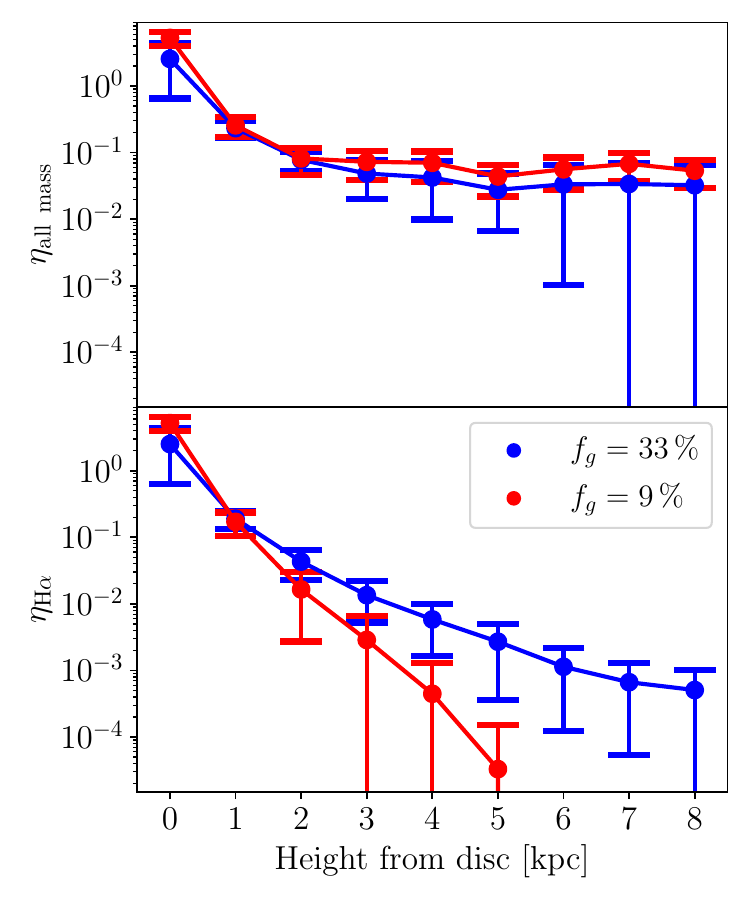}
    \caption{The mass loading factor $\eta$ as a function of the height from the disc. Top plot shows the mass loading of all the mass $\eta_{\rm\ all mass}$ and the bottom plot for the \Halpha traced (ionised) gas. The values are averaged from each patch at the relevant height bin and using several simulation outputs, which represents the spatial and temporal 1$\sigma$ errorbars.} 
    \label{fig:mass_loading_factor}
\end{figure}

We next evaluate the mass loading factor $\eta = \dot{M}_{\rm out}/{\rm SFR}$, which quantifies the amount of gas mass flowing out from the galaxy relative to its star formation rate. This is presented in Figure~\ref{fig:mass_loading_factor} as a function of height from the disc, for both the total and the ionised (${\rm H\alpha}$) gas. The mass outflow $\dot{M}_{\rm out}$ is calculated at each height bin (above and below the disc) by summing the amount of mass at that height that has a velocity away from the disc. The threshold for \Halpha gas, used with $\eta_{\rm H\alpha}$, was set as $\epsilon_{\rm H\alpha} = 10^{-30}\,{\rm erg\,s^{-1}\,cm^{-3}}$, same as mentioned in the previous section. At a distance of 2\,kpc from the disc plane, both simulations have a constant mean value $\eta\sim 0.04-0.07$ up to, at least, 8\,kpc, hinting towards a similar outflow history (averaged over time). To note is that the $f_{\rm g}=33\%$ galaxy has roughly 20 times the gas outflow strength $\dot{M}_{\rm out}$ as the $f_{\rm g}=9\%$ galaxy, but due to the difference in SFR being roughly the same ($39\Msolyr$ compared to $2\Msolyr$) they attain similar $\eta$ values. The simulations differ in terms of $\eta_{\rm H\alpha}$, which monotonically decreases with height in both galaxies, but the slope for the higher gas fraction is more shallow and thus has more extended \Halpha outflow relative its SFR.

We want to compare our results to observed $\eta$ for disc galaxies with similar stellar masses as our low/high gas fraction (proxy for low/high redshift) galaxies, ($4.10,\ 3.02)\,\cdot10^{10}\Msol$. Observations of both low and high-redshift disc galaxies with $M_*\sim 10^{10}\Msol$ using various gas tracers show a large range of possible mass loading factors $\eta\sim0.1-10$ \citep[e.g.][]{Chisholm+17, Swinbank+19, Fluetsch+19, Stuber+21, Xu+22, Weldon+24}{}{}. This large range could possible be narrowed by only comparing observations with the same galaxy characteristics (e.g. dwarf, starburst) and tracing the same gas phase (e.g. ionised, molecular). However, the accuracy of observations to determine $\eta$ is still under debate, due to variations in methodology, assumptions, and galaxy selection \citep[see e.g. the discussions in][]{Concas+22, Weldon+24}{}{}. In particular, \citet[][]{Concas+22}{}{} showed how the "velocity-subtracted" method can produce artificial broadening, which could be interpreted as outflow and cause a systematic overestimation of $\eta$ in observations of spatially unresolved sources \citep[see also][]{Genzel+14}{}{}.
% Dwarf starburst galaxies have yielded very varied results, with some studies suggesting little to no outflow \citep[negative feedback][]{}{}{} and others much higher outflows $\eta\gtrsim10$ \citep[due to a more shallow potential well][]{}{}{}.

Additionally, galaxies in large-scale cosmological simulations, such as {\small TNG50} \citep[][]{Nelson+19}{}{}, {\small EAGLE} \citep[][]{Mitchell+20}{}{}, and {\small FIRE-1} \& {\small FIRE-2} \citep[][]{Muratov+15, Pandya+21}{}{}, exhibit mass loading factors $\eta\sim1-10$ (but a few {\small FIRE-2} galaxies extending down to as low as $\eta\gtrsim0.1$) for stellar masses similar to our simulations. However, these simulations report the total mass outflow and are at odds with \Halpha observations of galaxies with $\eta\gtrsim0.001$ \citep[e.g.][]{HeckmanThompson17, Fluetsch+19, Swinbank+19, Concas+22, Marasco+23, Weldon+24}{}{}, which has been suggested to be due to other gas phases contributing more to the mass outflow than ionised gas. Our results agree with this idea, since $\eta_{\rm all\ mass}$ is roughly 100 times as large as $\eta_{\rm H\alpha}$ at 8\,kpc for the high gas fraction galaxy.  % However, this approach does not consider the mass in hot ($T\gtrsim10^6$\,K) ionised gas.
While the range of observed $\eta$ values is large, it is reassuring that our simulations achieve similar outflows to observations. However, the mass loading factor is still near the lower limit of what has been observed, which could be due to observational over-estimations of $\eta$ or restrictions of our model. For example, our simulations are not cosmological, meaning that any gas above the disc originated either from disc outflows or from the circumgalactic medium that was initiated with the galaxy. It is possible that inflows or minor mergers could contribute to the extraplanar gas by stripping more gas from the disc or allow for more shocked, warm, extraplanar gas. We do not include dust scattering, which has been found by RT models to contribute $\sim5-30\%$ \citep[][]{ WoodReynolds99, Barnes+15, Tacchella2022}{}{} of the total luminosity. Furthermore, our models do not include cosmic rays, magnetic fields or black hole feedback which could help drive outflows (and winds), increasing the \Halpha emission and turbulence outside of the disc. Additionally, \citet[][]{Rey+24} performed simulations of a dwarf galaxy and highlighted how tracing the diffuse gas at a higher spatial resolution better traces its propagation and can boost the outflow, resulting in a twice as large mass loading factor (at $r=5$\,kpc).%, but that the ISM remains largely unaffected.
% As RT post-processing would require notably more analysis, we leave dust scattering for future work. 

% , as outflows driven by star-formation is expected to reach $\eta\leq1$ \citep[see e.g.][]{Somerville+15}{}{}

% More massive or gas-rich disc galaxies with more bursty SFRs could possibly have more volatile outflows, resulting in higher $\eta$ and \Halpha emission above the disc. 
% \Halpha emission in extraplanar gas is rather low

% This supports that our simulations have realistic outflows and that the weak extraplanar \Halpha emission is indeed a property of these types of disc galaxies. The height at which observations probe gas outflows is not clear and one source of the large variation between observations might stem from the depths at which the observations and method, together, probe these outflows. 

% Going from 3 $\rightarrow$ 6 kpc in height decreases the ionised gas mass-loading factor by an order of magnitude in the high gas fraction case.

% Thus, another mechanism is needed to explain the lack of \Halpha emission above/below the disc.

%%%%%%%%%%%%%%%%%%%%%%%%%%%%%%%%%%%%%%%%%%%%%%%%%%%%%%%%%%%%%%%%%%%%%%%%%%%%%%%%%%%%%%%%%%%%%%
\section{Conclusions}
The \Halpha emission line is a versatile observational tool for deriving various properties in both low- and high-redshift galaxies. However, \Halpha is emitted by both dense and diffuse gas, which contribute differently to the \Halpha emission line profile. Furthermore, diffuse, \Halphaalt-bright gas is ubiquitous inside and around the galactic disc. Untangling the various contributions to the \Halpha line is important for interpreting the underlying, complex physical processes.

In this paper, we present radiation hydrodynamic simulations of isolated disc galaxies in order to evaluate the strength and kinematics of \Halpha emitting gas in outflows, diffuse gas regions, and dense gas regions. We model Milky Way-like galaxies, in terms of mass and size, with different gas fractions ($f_{\rm g}=9\%$ and 33\%) as a proxy of the galaxy at different epochs of its evolution. We carry out mock observation of the \Halpha emission line, which, after careful numerical treatment of unresolved \HII regions (Appendix~\ref{appendix:stromgren}), match empirical \LHaalt-SFR relations \citep[][]{Kennicutt98}{}{}.

The key findings in this work are summarised below:
\begin{enumerate}
    \item We highlight a distinct difference in the \Halpha luminosity and kinematic signature between the diffuse and dense gas emitting strongly in \Halphaalt, as is also found between observations of DIG and dense \HII region gas \citep[see e.g.][and references therein]{Law+22}{}{}. In the gas rich galaxy ($f_{\rm g}=33\%$), the dense gas reaches at most \sigmaHa$\lesssim 40$\kmsec, with a mean value of 29\kmsec, while the diffuse gas reaches $\sigma_{\rm H\alpha} \lesssim 80\,{\rm km\, s^{-1}}$, with a mean value of 54\kmsec. 
    \item The evolution of \sigmaHa with gas fraction indicates that our Milky Way-like galaxy was more turbulent when it was younger. The higher gas fraction galaxy is roughly twice as turbulent as the low gas fraction galaxy. This \sigmaHa evolution is the same for \Halpha emission from \HII regions, but is more shallow for \Halpha from DIG.
    %The low gas fraction galaxy has an \Halpha emission line (ionised gas) roughly half as broad (turbulent) when looking at the total \Halpha emission. This is also true for \Halpha emission from dense, \HII regions. However, the \Halpha emission in the DIG has a more shallow evolution. 
    \item The contribution of the diffuse ($\rho_{\rm H}\lesssim 0.1\,{\rm cm^{-3}}$) gas to \LHa is $\lesssim 10\,\%$. This matches with the lower-limits found by observations of the DIG (see the discussion in Section~\ref{sec:intro}). However, several observational studies suggests the DIG can constitute the majority of \LHaalt and some numerical work with similar setup to ours, \citet[][]{Tacchella2022}{}{}, agrees with this high DIG fraction. The major difference between our and their setup is that we do not perform post-processing to trace the \Halpha emission line. Our study encourages further work into post-processing of \Halpha emission line propagation in RHD simulations, to capture the impact of dust scattering and absorption on the observed \Halpha line profile.
    \item Gas outside of the galactic disc has a negligible to the total \Halpha luminosity, despite that the mass outflow in our high gas-fraction galaxy match with observed values of galaxies with similar stellar mass. Although, extraplanar gas contributes 45\% of the \Halpha line profile in the extended wings ($v_z\geq200$\kmsec). Thus, the overall galaxy disc dynamics could be well-constrained with observations of the \Halpha line; taking into consideration any kinematic offset between the DIG and \HII regions (see points above).
    \item The \sigmaHa of gas that exists several kpc above the disc is much more turbulent compared to the mid-plane, increasing by a factor of $\sim2$, depending on the simulation. However, the luminosity at these heights corresponds to only a few percent of the total \LHaalt, which has an \Halpha surface density below the sensitivity limit of {\small MUSE} in local galaxies and JWST at redshift $\sim1.5$ (see Section~\ref{sec:Ha_extraplanar}). This suggests that the \Halpha emission can be readily observed a few kpc above the disc and the predictions of our simulations on the \Halpha vertical structure can be tested.
\end{enumerate}

\section*{Acknowledgements}
The computations and data storage were enabled by resources provided by LUNARC - The Centre for Scientific and Technical Computing at Lund University. OA acknowledges support from the Knut and Alice Wallenberg Foundation, and from the Swedish Research Council (grant 2019-04659). MPR is supported by the Beecroft Fellowship funded by Adrian Beecroft.
FR acknowledges support provided by the University of Strasbourg Institute for Advanced Study (USIAS), within the French national programme Investment for the Future (Excellence Initiative) IdEx-Unistra. 

Simulation outputs were analysed using tools from the {\small YT} project \citep[][]{Turk+11}{}{}, numpy \citep[][]{Harris+20}{}{}, and {\small MATPLOTLIB} for {\small PYTHON} \citep[][]{Hunter+07}{}{}.

%%%%%%%%%%%%%%%%%%%%%%%%%%%%%%%%%%%%%%%%%%%%%%%%%%
\section*{Data Availability}
The data underlying this article will be shared on reasonable request to the corresponding author.


\bibliographystyle{mnras}
%\bibliography{complete.bib} % references.bbl
\input{main.bbl} % references.bbl

%%%%%%%%%%%%%%%%%%%%%%%%%%%%%%%%%%%%%%%%%%%%%%%%%%

%%%%%%%%%%%%%%%%% APPENDICES %%%%%%%%%%%%%%%%%%%%%

\appendix

\section{The impact of unresolved H{\small II} regions on H$\alpha$ parameters}\label{appendix:stromgren}

\citet[][]{Deng+23}{}{} showed that to properly resolve the \HII regions in numerical simulations, the Str{\"o}mgren radius needs to be resolved by 10-100 simulation cells. Not resolving the \HII regions causes these cells to be heated up by the ionising radiation from the bright stars within, which results in an unrealistic gas mix of partially-ionised, warm ($T\sim8\,000$\,K), dense ($\rho\geq100\,{\rm cm^{-3}}$) gas. These artificial conditions are prime for \Halpha emission and can contaminate analysis of \Halpha quantities. Here we evaluate the impact of unresolved \HII regions on the \Halpha emission and kinematics computed in this paper. 

The size of an idealised \HII region can approximated by the Str{\"o}mgren radius. The equation describes the size of a spherical ionised bubble formed as one or more very young and luminous stars ionise the surrounding gas, and can be written as
\begin{align}
    R_{\rm SG} = \bigg({\frac{3 Q_0}{4\pi n_{\rm H}^2 \alpha_{\rm B}  }}\bigg)^{1/3}.
\end{align}
Here, $Q_0$ is the rate of ionising photons emitted by the stars, $n_{\rm H}$ is the hydrogen density, and $\alpha_{\rm B}$ is the effective hydrogen recombination rate, which can be described as a function of gas temperature 
\begin{align}
    \alpha_{\rm B}(T) = 2.753\cdot 10^{-14}\, \bigg(\frac{315614}{T}\bigg)^{1.5} \Bigg(1+\bigg(\frac{315614}{T}\bigg)^{0.407}\Bigg)^{-2.42}.
\end{align}
This expression is a fit made by \citet[][]{HuiGnedin97}{}{}, using observational data from \citet[][]{Ferland+92}{}{}.

\begin{figure}
\includegraphics[width=0.45\textwidth]{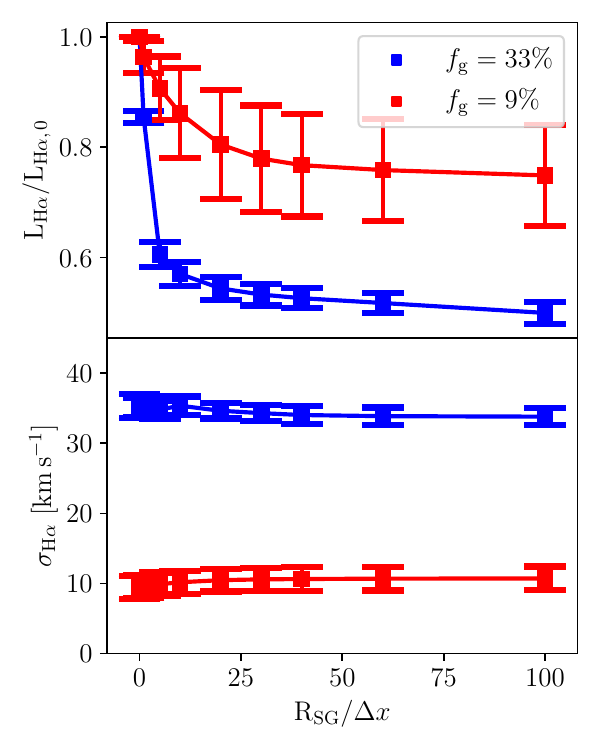}
    \caption{The \Halpha luminosity and velocity dispersion above different criteria for resolving the Str{\"o}mgren sphere $R_{\rm SG}$ with cells of size $\Delta x$. Here, ${\rm L_{H\alpha,\,0}}$ is the un-filtered luminosity of each simulations, which is $5.5\cdot10^{41}\,{\rm erg\,s^{-1}}$ and $8.5\cdot10^{42}\,{\rm erg\,s^{-1}}$ for \texttt{fg10} and \texttt{fg50}, respectively. }
    \label{fig:R_SG}
\end{figure}

We implement a criteria for the number of resolved Str{\"o}mgren radii ($R_{\rm SG}/\Delta x$) in our simulations by reducing the \Halpha emissivity to the mean value of resolved \HII regions in our galaxies ($\epsilon_{\rm H\alpha}=2\cdot10^{-20} {\rm erg\,s^{-1}\,cm^{-3}}$ for \texttt{fg50} and $\epsilon_{\rm H\alpha}=7\cdot10^{-21} {\rm erg\,s^{-1}\,cm^{-3}}$ for \texttt{fg10}) for the regions unresolved by a certain amount of cells. Figure~\ref{fig:R_SG} shows \LHa and \sigmaHa as a function of how many cells $R_{\rm SG}$ is resolved. As can be seen, there is a strong decrease in the \Halpha luminosity at 1-10 cells resolved, as most of the highly luminous, and unresolved, regions are filtered away. When $R_{\rm SG}/\Delta x \approx 10$ unresolved cells are filtered, the decrease in luminosity is much more shallow, indicating that the most problematic cells have been corrected. The expected luminosity from the \LHaalt-SFR relation \citep[][]{Kennicutt98}{}{} for these SFRs is around $2.5\cdot10^{41} {\rm erg\,s^{-1}}$ and $5\cdot10^{42} {\rm erg\,s^{-1}}$ for the low- and high-gas fraction simulations, respectively. All of this indicates that by filtering cells that do not resolve the Str{\"o}mgren sphere with at least 10 cells, we correct for most of the unresolved \HII regions and achieve a much better match with \LHa predicted.

Furthermore, the lower plot in this figure suggests that the \Halpha velocity dispersion for the entire galaxy remains largely unaffected by this filtering process. However, this \sigmaHa is integrated over the entire galaxy (comparable to observations of unresolved galaxies), rather than computing \sigmaHa in each grid patch and then calculating a mean value from that, which is the common approach for resolved galaxy observations. Therefore, it is not directly comparable to \sigmaHa we compute in the main body of the text, but, as discussed in Section~\ref{sec:DIGvsHII}, \sigmaHa remains largely unaffected by altering the grid patch size.

%%%%%%%%%%%%%%%%%%%%%%%%%%%%%%%%%%%%%%%%%%%%%%%%%%

% Don't change these lines
\bsp	% typesetting comment
\label{lastpage}
\end{document}